\documentclass[11pt]{article}
\pdfoutput=1
\usepackage{fullpage}
\usepackage{tikz-cd}
\usepackage{cite}
\usepackage{graphicx}
\usepackage{amsmath}
\usepackage{amssymb}
\usepackage{subcaption}
\title{}
\date{}
\renewcommand{\vec}[1]{\mbox{\boldmath$ #1 $}}

\def\beq{\begin{equation}}
\def\eeq{\end{equation}}
\allowdisplaybreaks
\begin{document}
\bibliographystyle{utphys}

\newcommand\n[1]{\textcolor{red}{(#1)}} 
\newcommand{\diff}{\mathop{}\!\mathrm{d}}
\newcommand{\lb}{\left}
\newcommand{\rb}{\right}
\newcommand{\f}{\frac}
\newcommand{\pd}{\partial}
\newcommand{\tr}{\text{tr}}
\newcommand{\fdiff}{\mathcal{D}}
\newcommand{\im}{\text{im}}
\let\caron\v
\renewcommand{\v}{\mathbf}
\newcommand{\T}{\tensor}
\newcommand{\R}{\mathbb{R}}
\newcommand{\C}{\mathbb{C}}
\newcommand{\Z}{\mathbb{Z}}
\newcommand{\msbar}{\ensuremath{\overline{\text{MS}}}}
\newcommand{\DIS}{\ensuremath{\text{DIS}}}
\newcommand{\abar}{\ensuremath{\bar{\alpha}_S}}
\newcommand{\bb}{\ensuremath{\bar{\beta}_0}}
\newcommand{\rc}{\ensuremath{r_{\text{cut}}}}
\newcommand{\Nd}{\ensuremath{N_{\text{d.o.f.}}}}
\setlength{\parindent}{0pt}

\titlepage
\begin{flushright}
QMUL-PH-20-08\\
\end{flushright}

\vspace*{0.5cm}

\begin{center}
{\bf \Large Topology and Wilson lines: global aspects of the double
  copy}

\vspace*{1cm} 
\textsc{Luigi Alfonsi\footnote{l.alfonsi@qmul.ac.uk},
  Chris D. White\footnote{christopher.white@qmul.ac.uk}, 
 and Sam Wikeley\footnote{s.wikeley@qmul.ac.uk}} \\

\vspace*{0.5cm} Centre for Research in String Theory, School of
Physics and Astronomy, \\
Queen Mary University of London, 327 Mile End
Road, London E1 4NS, UK\\

\end{center}

\vspace*{0.5cm}

\begin{abstract}
The Kerr-Schild double copy relates exact solutions of gauge and
gravity theories. In all previous examples, the gravity solution is
associated with an abelian-like gauge theory object, which linearises
the Yang-Mills equations. This appears to be at odds with the double
copy for scattering amplitudes, in which the non-abelian nature of the
gauge theory plays a crucial role. Furthermore, it is not yet clear
whether or not global properties of classical fields - such as
non-trivial topology - can be matched between gauge and gravity
theories. In this paper, we clarify these issues by explicitly
demonstrating how magnetic monopoles associated with arbitrary gauge
groups can be double copied to the same solution (the pure NUT metric)
in gravity. We further describe how to match up topological
information on both sides of the double copy correspondence,
independently of the nature of the gauge group. This information is
neatly expressed in terms of Wilson line operators, and we argue
through specific examples that they provide a useful bridge between
the classical double copy and the BCJ double copy for scattering
amplitudes.
\end{abstract}

\vspace*{0.5cm}

\section{Introduction}
\label{sec:intro}

Since its inception just over a decade
ago~\cite{Bern:2008qj,Bern:2010ue,Bern:2010yg}, the double copy has
remained an intensively-studied relationship between gauge and gravity
theories. Its first application was to scattering amplitudes, and
there is by now a large amount of evidence for its validity at
arbitrary loop
orders~\cite{Bern:2010ue,Bern:1998ug,Green:1982sw,Bern:1997nh,Carrasco:2011mn,Carrasco:2012ca,Mafra:2012kh,Boels:2013bi,Bjerrum-Bohr:2013iza,Bern:2013yya,Bern:2013qca,Nohle:2013bfa,
  Bern:2013uka,Naculich:2013xa,Du:2014uua,Mafra:2014gja,Bern:2014sna,
  Mafra:2015mja,He:2015wgf,Bern:2015ooa,
  Mogull:2015adi,Chiodaroli:2015rdg,Bern:2017ucb,Johansson:2015oia,Oxburgh:2012zr,White:2011yy,Melville:2013qca,Luna:2016idw,Saotome:2012vy,Vera:2012ds,Johansson:2013nsa,Johansson:2013aca,Bargheer:2012gv,Huang:2012wr,Chen:2013fya,Chiodaroli:2013upa,Johansson:2014zca,Johansson:2017srf,Chiodaroli:2017ehv,Chen:2019ywi,Cheung:2020uts},
either with or without supersymmetry. The double copy has also been
extended to classical
solutions~\cite{Monteiro:2014cda,Luna:2015paa,Luna:2016due,Goldberger:2016iau,Anastasiou:2014qba,Borsten:2015pla,Anastasiou:2016csv,Anastasiou:2017nsz,Cardoso:2016ngt,Borsten:2017jpt,Anastasiou:2017taf,Anastasiou:2018rdx,LopesCardoso:2018xes,Goldberger:2017frp,Goldberger:2017vcg,Goldberger:2017ogt,Luna:2016hge,Luna:2017dtq,Shen:2018ebu,Levi:2018nxp,Plefka:2018dpa,Cheung:2018wkq,Carrillo-Gonzalez:2018pjk,Monteiro:2018xev,Plefka:2019hmz,Maybee:2019jus,Johansson:2019dnu,PV:2019uuv,Carrillo-Gonzalez:2019aao,Bautista:2019evw,Moynihan:2019bor,Bah:2019sda,CarrilloGonzalez:2019gof,Goldberger:2019xef,Kim:2019jwm,Banerjee:2019saj,Moynihan:2020gxj,Luna:2018dpt},
some of which are exact solutions of their respective field
equations. There are ongoing hopes that this may provide new
calculational tools for classical General Relativity, including for
astrophysical applications such as gravitational wave detection. In
parallel to this effort, however, it is worth examining the conceptual
underpinnings of the double copy correspondence, and to try to
ascertain the limits (if any) of its validity. If, for example, the
double copy can be extended to arbitrary (non)-perturbative solutions
in gauge and gravity theories, this clearly indicates a profound new
correspondence between different quantum field theories, that may even
suggest that our traditional approach to thinking about field theory needs
revisiting. One way to probe this is to find explicit non-perturbative
solutions in theories that enter the double copy correspondence, such
as the biadjoint scalar theory considered in
refs.~\cite{White:2016jzc,DeSmet:2017rve,Bahjat-Abbas:2018vgo,Bahjat-Abbas:2020cyb}. \\

There is, however, another approach to extending the conceptual
framework of the double copy. This is to note that nearly all
previous incarnations in either a classical or quantum context have
involved local quantities. Scattering amplitudes, for example, have
locality built in, even if this is not manifest. Furthermore, the
classical double copy initiated in ref.~\cite{Monteiro:2014cda}
involves products of fields at the same spacetime point, and thus is
again manifestly local. If it is true that there is a deep connection
between gauge and gravity theories, it must be possible to match up
{\it global} data - such as topological information - between them. This was examined recently in
ref.~\cite{Berman:2018hwd} which considered the single copy of the
Eguchi-Hanson instanton, alas without providing any conclusive
evidence that topological information can be
double-copied. Nevertheless, the idea clearly deserves further
investigation, and is thus the main aim of this paper.\\

\begin{figure}
\begin{center}
\scalebox{0.6}{\includegraphics{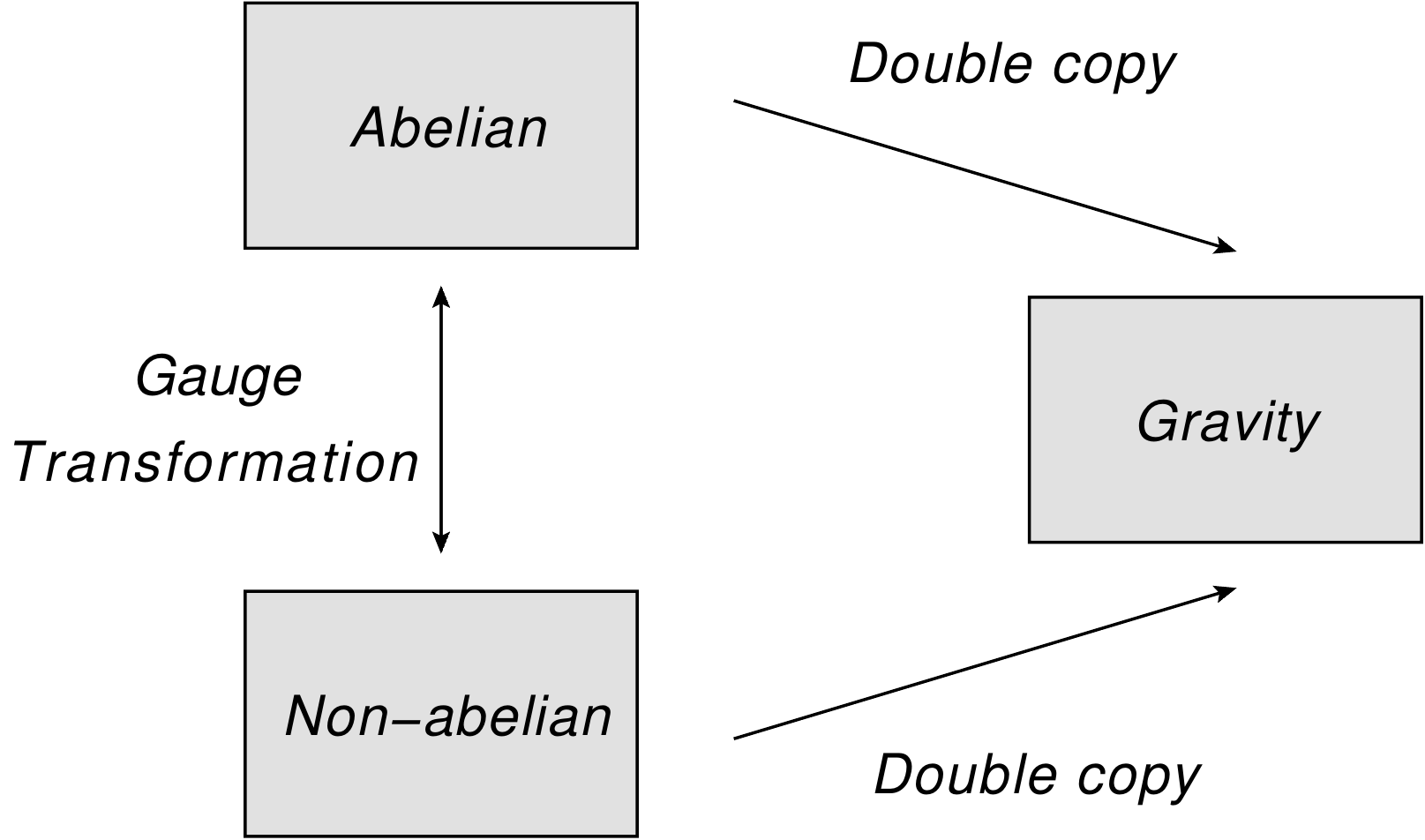}}
\caption{Conceptual structure of the double copy for exact solutions, in which (non-)abelian objects must double copy to the {\it same} gravity solution, regardless of the gauge group.}
\label{fig:gaugecopy}
\end{center}
\end{figure}

For scattering amplitudes and (non-linear) perturbative classical
solutions, the non-abelian nature of the gauge theory plays a crucial
role in the formulation of the double copy procedure. As is by now
well-known, there exists a special class of gravity solutions -
time-independent Kerr-Schild metrics - whose single copy gauge
solution obeys the {\it linearised} Yang-Mills equations, and thus
looks like an abelian gauge theory
solution~\cite{Monteiro:2014cda}. In what sense the gauge theory
solution is genuinely non-abelian is then rather obscure. However, by
analogy with amplitudes we must be able to find a construction such
that exact classical solutions with different gauge groups map to the
{\it same} gravity solution, as shown in figure~\ref{fig:gaugecopy}.
This follows from the fact that colour information is removed when
performing the double copy, so that a given gravity solution does not
care which gauge group one started with. Indeed,
figure~\ref{fig:gaugecopy} has a precedent in the study of infrared
singularities of amplitudes~\cite{Oxburgh:2012zr}, and was also
considered for solutions of SU(2) gauge theory in
ref.~\cite{Bahjat-Abbas:2020cyb} (see
refs.~\cite{Bargheer:2012gv,Huang:2012wr} for yet more
examples). However, this would appear to forbid the ability to map
topological information between gauge and gravity solutions:
topological invariants (e.g. characteristic classes) of a gauge field
usually rely crucially on which particular gauge group one considers,
and this information is irrelevant after taking the double copy.\\

In this paper, we will see that it is indeed possible to relate exact
solutions with non-trivial topology between gauge and gravity theories
in a meaningful way. First, we will take
a particular solution of pure gauge theory, namely a (singular)
point-like magnetic monopole. For general gauge groups, this is known
to have a non-trivial topology, where the latter may or may not be
classified by a known topological invariant. However, it can always be
characterised by a certain patching condition between gauge fields in
different spatial regions. Furthermore, the non-abelian monopole can always be written in a gauge in which it takes the form of a trivially dressed
abelian-like monopole solution, whose double copy is already
known~\cite{Luna:2015paa} to be the Taub-NUT solution~\cite{Taub,NUT}. There is an established patching condition for the Taub-NUT metric
that characterises its non-trivial topology~\cite{Bossard:2008sw}, and
we will see that this precisely corresponds to the similar condition
in gauge theory, independently of the nature of the gauge group. Thus,
we provide an explicit example of an exact solution whose local and
global properties obey the construction of figure~\ref{fig:gaugecopy},
and which fully generalises the preliminary observations regarding
SU(2) monopoles made recently in ref.~\cite{Bahjat-Abbas:2020cyb}.
Furthermore, we will see that all of our patching conditions - in
either gauge or gravity theories - can be expressed in terms of
certain Wilson line operators. This will allow us to make contact
between the study of magnetic monopoles in this paper, and previous
results concerning the structure of scattering amplitudes in
special kinematic limits~\cite{Oxburgh:2012zr,Melville:2013qca}. \\

The structure of this paper is as follows. In section~\ref{sec:review},
we briefly review the Kerr-Schild double copy. In
section~\ref{sec:topology}, we examine (non-)abelian monopoles, and describe the characterisation of their non-trivial topology. In
section~\ref{sec:gravity}, we study the topology of
Taub-NUT spacetime, and relate this to the single copy gauge theory
solutions considered in section~\ref{sec:topology}. The
relevance and implications of Wilson lines will be outlined in
section~\ref{sec:Wilson}. Finally, we discuss our results and conclude
in section~\ref{sec:conclude}.

\section{The Kerr-Schild double copy}
\label{sec:review}

The aim of the classical double copy is to associate a given
non-abelian gauge theory solution with a gravitational counterpart, in
a way that overlaps with the known BCJ double copy for scattering
amplitudes where
appropriate~\cite{Bern:2008qj,Bern:2010ue,Bern:2010yg}. For general
solutions, this must be carried out order-by-order in the relevant
coupling constants. However, a certain special family of exact gravity
solutions is known, which all have well-defined gauge theory single
copies. These are the Kerr-Schild metrics (see
e.g.~\cite{Stephani:2003tm} for a modern review), which can be defined
as follows~\footnote{Note that we use the $(+,-,-,-)$ metric signature
  throughout.}:
\begin{equation}
g_{\mu\nu}=\eta_{\mu\nu}+\kappa h_{\mu\nu},\quad h_{\mu\nu}=\phi 
k_\mu k_\nu.
\label{KSdef}
\end{equation}
Here $h_{\mu\nu}$ is the graviton field representing the deviation
from the Minkowski metric $\eta_{\mu\nu}$, and $\kappa=\sqrt{32\pi
  G_N}$, where $G_N$ is the Newton constant. The Kerr-Schild form
corresponds to the graviton decomposing into a scalar field $\phi$
multiplying an outer product of a vector $k_\mu$ with itself, where
the latter must satisfy null and geodesic properties:
\begin{equation}
\eta_{\mu\nu} k^\mu k^\nu=g_{\mu\nu} k^\mu k^\nu=0,\quad
k\cdot \partial k^\mu=0.
\label{KSconditions}
\end{equation}
This ansatz turns out to linearise the Einstein equations, such that
exact solutions can be more easily found. Each such solution is
characterised by the explicit forms of $\phi$ and $k_\mu$, and
ref.~\cite{Monteiro:2014cda} proved that for every
time-independent Kerr-Schild graviton, one may then construct a single copy
gauge field
\begin{equation}
{\bf A}_\mu=(c^a {\bf T}^a)\phi k_\mu,
\label{singlecopy}
\end{equation}
which obeys the linearised Yang-Mills equations. Here $c^a$ is an
arbitrary constant colour vector, and ${\bf T}^a$ is a generator of the
gauge group. The extension to symmetries other than time translational
invariance has been considered in
ref.~\cite{Carrillo-Gonzalez:2017iyj}. Clearly eq.~(\ref{singlecopy})
is not unique: for a particular gauge group, any choice of $c^a$ will
suffice. We may furthermore pick any gauge group. Indeed, it is
instructive to write eq.~(\ref{singlecopy}) as
\begin{equation}
{\bf A}_\mu=(c^a {\bf T}^a)A_\mu^{\rm abel.},
\label{singlecopy2}
\end{equation}
where $A_\mu^{\rm abel.}$ is the solution of a purely abelian gauge
theory. If the latter has a known Kerr-Schild double copy, any
solution of the form of eq.~(\ref{singlecopy2}), in which the colour
structure completely factorises, can also be easily double copied. One
simply strips off the colour charge $(c^a {\bf T}^a)$, and double
copies the gauge field as if it were abelian. Note that this provides
an explicit realisation of figure~\ref{fig:gaugecopy}, in which both
abelian and non-abelian classical solutions are taken to map to the
same gravity solution.\\

There is a growing list of specific cases of the Kerr-Schild double
copy which have been examined in detail. Arguably the simplest is that
of a point-like mass $M$ in gravity giving rise to the Schwarzschild
metric, for which the single copy is a point-like (electric)
charge~\cite{Monteiro:2014cda}~\footnote{Going the other way, the
  double copy of a point charge may be more general than a pure
  gravity solution, as discussed in
  refs.~\cite{Goldberger:2017frp,Luna:2016hge,Kim:2019jwm}.}. Particularly
relevant for this paper is the Taub-NUT solution in
gravity~\cite{Taub,NUT}, which was examined from a double copy point
of view in ref.~\cite{Luna:2015paa}. This has a Schwarzschild-like
mass term, but in addition a so-called {\it NUT charge} $N$, which
gives rise to a rotational character in the gravitational field at
infinity. In particular (Plebanski) coordinates, the relevant graviton
field can be written in a so-called double Kerr-Schild
form~\cite{Chong:2004hw}:
\begin{equation}
h_{\mu\nu}=M\phi k_\mu k_\nu+N\psi l_\mu l_\nu,
\label{doubleKS} 
\end{equation}
where $\phi$ and $\psi$ are distinct scalar fields, and $k_\mu$,
$l_\mu$ different vectors that obey certain mutual orthogonality
conditions. Such an ansatz will not linearise the Einstein equations
in general, but happens to do so for Taub-NUT. One may then single
copy the solution to create a gauge field
\begin{equation}
{\bf A}_\mu=(c^a {\bf T}^a)\phi k_\mu+
(\tilde{c}^a {\bf T}^a)\psi l_\mu.
\label{ANUT}
\end{equation}
Reference~\cite{Luna:2015paa} analysed this solution in an abelian
gauge theory, and concluded that it was a dyon, possessing both
electric and magnetic monopole charge. These are represented by the
first and second terms in eq.~(\ref{ANUT}) respectively, such that for
the case of a pure NUT charge in the gravity theory, one may follow
eq.~(\ref{singlecopy2}) and write 
\begin{equation}
{\bf A}_\mu=(\tilde{c}^a {\bf T}^a)A_\mu^{\rm D},
\label{singlecopy3}
\end{equation}
where~\footnote{Here, and for the rest of the paper, spherical coordinates are defined such that 
\begin{equation*}
  A_xdx + A_ydy + A_zdz = A_rdr + A_{\theta}d\theta + A_{\phi}d\phi.
\end{equation*}}
\begin{equation}
A_\mu^D=-\tilde{g}\left(0,0,0,\cos{\theta}-1\right)
\label{AmuD}
\end{equation}
is the well-known gauge potential for a Dirac magnetic monopole in
the U(1) (abelian) gauge theory of electromagnetism. We have here
expressed the result in spherical polar coordinates
$(t,r,\theta,\varphi)$, and denoted the magnetic charge parameter that
would arise in a purely abelian theory by
$\tilde{g}$. Reference~\cite{Luna:2015paa} did not explicitly check
that eq.~(\ref{singlecopy3}) represents a genuine magnetic monopole
solution in the case of a non-abelian gauge theory. We will see in the
following section that this is indeed the case, such that $\tilde{g}
(\tilde{c}^a{\bf T}^a)$ is the appropriate non-abelian generalisation
of the magnetic charge. In line with the above comments, the double
copy of eq.~(\ref{singlecopy3}) is straightforward, and always results
in a pure NUT charge in gravity.

\section{(Non)-abelian magnetic monopoles and their topology}
\label{sec:topology}

In the previous section, we have seen that (non)-abelian magnetic
monopoles with arbitrary gauge groups all map to a pure NUT charge in
gravity. In this section, we discuss how to characterise the
non-trivial topology of monopole solutions, in both abelian and
non-abelian gauge theories. 

\subsection{Abelian case}
\label{sec:abelian}

The gauge potential for the Dirac monopole of eq.~(\ref{AmuD}) has a
singularity at $\theta=\pi$, corresponding to the entire negative
$z$-axis, which is usually referred to as the {\it Dirac
  string}. Whilst it is not possible to remove the physical singularity at $r=0$
(i.e. the location of the singular point-like monopole), the orientation of the string singularity can be modified by performing a gauge transformation. To construct a
gauge field that is everywhere non-singular away from the origin, one must employ at least two coordinate patches, such that the gauge
field is non-singular in each. Where these coordinate patches overlap the gauge fields defined in each can then be glued together.\\

This idea was first discussed by Wu and Yang, who gave an elegant
fibre bundle interpretation~\cite{Wu:1975es}. To briefly summarise for
the uninitiated: gauge fields can be defined as connections on {\it
  principal fibre bundles}, namely manifolds that look locally like a
product space ${\cal M}\times G$, where ${\cal M}$ denotes Minkowski
spacetime, and $G$ the gauge group. One refers to ${\cal M}$ as the
{\it base space} and $G$ as the {\it fibre}, where there is one fibre
attached to each point in the base space. Whilst such a bundle may
look locally like a product space, it might differ from this
globally. Thus, for a given gauge group, there may be more than one fibre
bundle with the same local structure. The classification of
topologically non-trivial gauge fields is then entirely equivalent to
the classification of fibre bundles. \\

\begin{figure}
\begin{center}
\scalebox{0.6}{\includegraphics{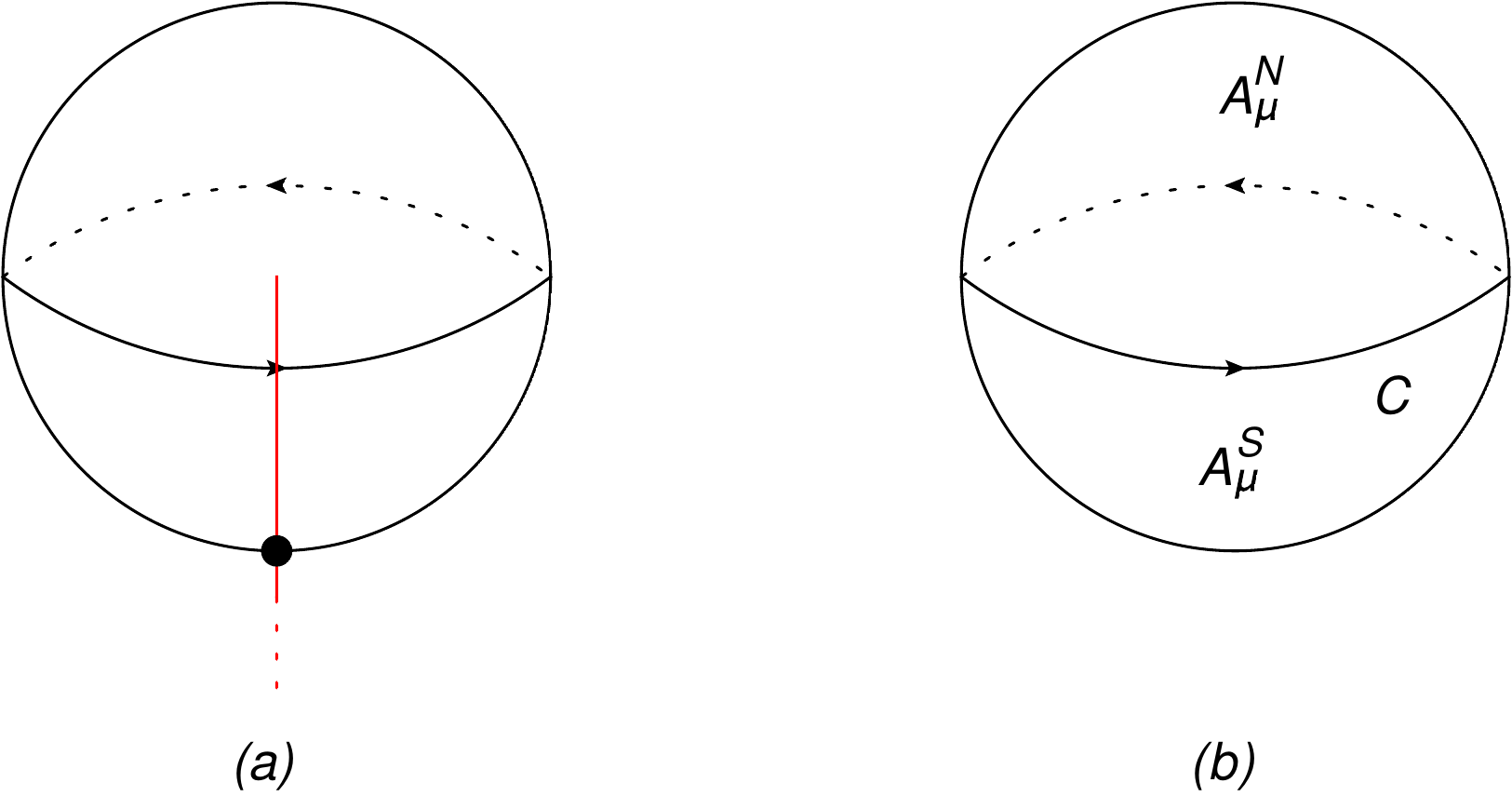}}
\caption{(a) The Dirac monopole field of eq.~(\ref{AmuD}) has a
  string-like singularity that cuts the sphere at infinity at the
  south pole; (b) Wu-Yang construction of a non-singular gauge field:
  one may take non-singular gauge fields in the northern and southern
  hemispheres, such that they are patched together by a gauge
  transformation on the equator $C$.}
\label{fig:patching}
\end{center}
\end{figure}

The Dirac monopole is singular at the origin and thus the base space is Minkowski spacetime with the origin
removed, ${\cal M}-\{0\}$. The spatial part of this manifold
can be continuously deformed into the surface of a sphere at
infinity. Adding the time direction, the base space is now
topologically equivalent to $S^2\times {\cal R}$. The monopole field
of eq.~(\ref{AmuD}) is tangential to the surface of the sphere, and it
is well known that it is impossible to define a vector field on $S^2$
without it being ill-defined at one point. This itself explains the
presence of the Dirac string, which may be taken to puncture the
sphere at infinity at the singular point, as shown in
figure~\ref{fig:patching}. \\

We indeed then need at least two coordinate patches in order to define
the gauge field throughout all space, which we can take to be the
northern and southern hemispheres respectively. We are free to define
separate gauge fields $A_\mu^N$ and $A_\mu^S$ in each patch, as shown
in figure~\ref{fig:patching}, provided that they are related by a
gauge transformation in the region of overlap, namely on the equator $C$. This is straightforward to achieve in
practice. Firstly, the field of eq.~(\ref{AmuD}) is already
non-singular in the northern hemisphere, and so can be taken as
$A_\mu^N$. One may then define the gauge field 
\begin{equation}
A_\mu^S=-\tilde{g}\left(0,0,0,\cos{\theta}+1\right).
\label{AmuS}
\end{equation}
This is related to $A_\mu^N$ by the gauge transformation 
\begin{equation}
A_\mu^S=A_\mu^N-\frac{i}{g}S(\varphi)\partial_\mu S^{-1}(\varphi),
\label{AmuStrans}
\end{equation}
where $g$ is the coupling constant, and 
\begin{equation}
S(\varphi)=e^{2ig\tilde{g}\varphi}
\label{Sdef}
\end{equation}
is an element of the gauge group. The field of eq.~(\ref{AmuS}) has a
string-like singularity on the positive $z$ axis, and is thus
non-singular throughout the southern hemisphere as
required. Furthermore, the two fields can be matched at the equator,
where they are equivalent up to the gauge transformation in
eq.~(\ref{AmuStrans}). This completes the construction of a
single-valued, non-singular gauge field away from the origin.\\

The patching together of two distinct gauge fields constitutes a
non-trivial topology, which turns out to be related to the magnetic
charge. To see this, note that the magnetic flux is given by
\begin{equation}
\Phi_B=\iint_S F_{\mu\nu}d\Sigma^{\mu\nu},
\label{magflux}
\end{equation}
where $d\Sigma^{\mu\nu}$ is the area element on the surface $S$
corresponding to the 2-sphere at infinity. Separating the surface
integral into two separate contributions for the northern and southern
hemispheres, one may apply Stokes' theorem to rewrite
eq.~(\ref{magflux}) as
\begin{align}
\Phi_B&=\oint_C dx^\mu \left(A_\mu^N-A_\mu^S\right)=4\pi \tilde{g},
\label{magflux2}
\end{align}
where we have used eqs.~(\ref{AmuStrans}, \ref{Sdef}). This justifies
the statement that $\tilde{g}$ represents the amount of magnetic
charge. Furthermore, the requirement that the gauge transformation of
eq.~(\ref{AmuStrans}) be single-valued (i.e. that $S(2\pi)=S(0)$ in
eq.~(\ref{Sdef})) implies the famous {\it Dirac quantisation
  condition}
\begin{equation}
g\tilde{g}=\frac{n}{2},\quad n\in{\mathbb Z},
\label{Diracquant}
\end{equation}
which relates the electric and magnetic charge. The relation to the
topology of the fibre bundle can be seen from the fact that there are
discretely different U(1) principle bundles, each of which is
classified by the so-called {\it first Chern number}. The latter, up
to a constant factor, is given precisely by eq.~(\ref{magflux}). \\

\begin{figure}
\begin{center}
\scalebox{0.7}{\includegraphics{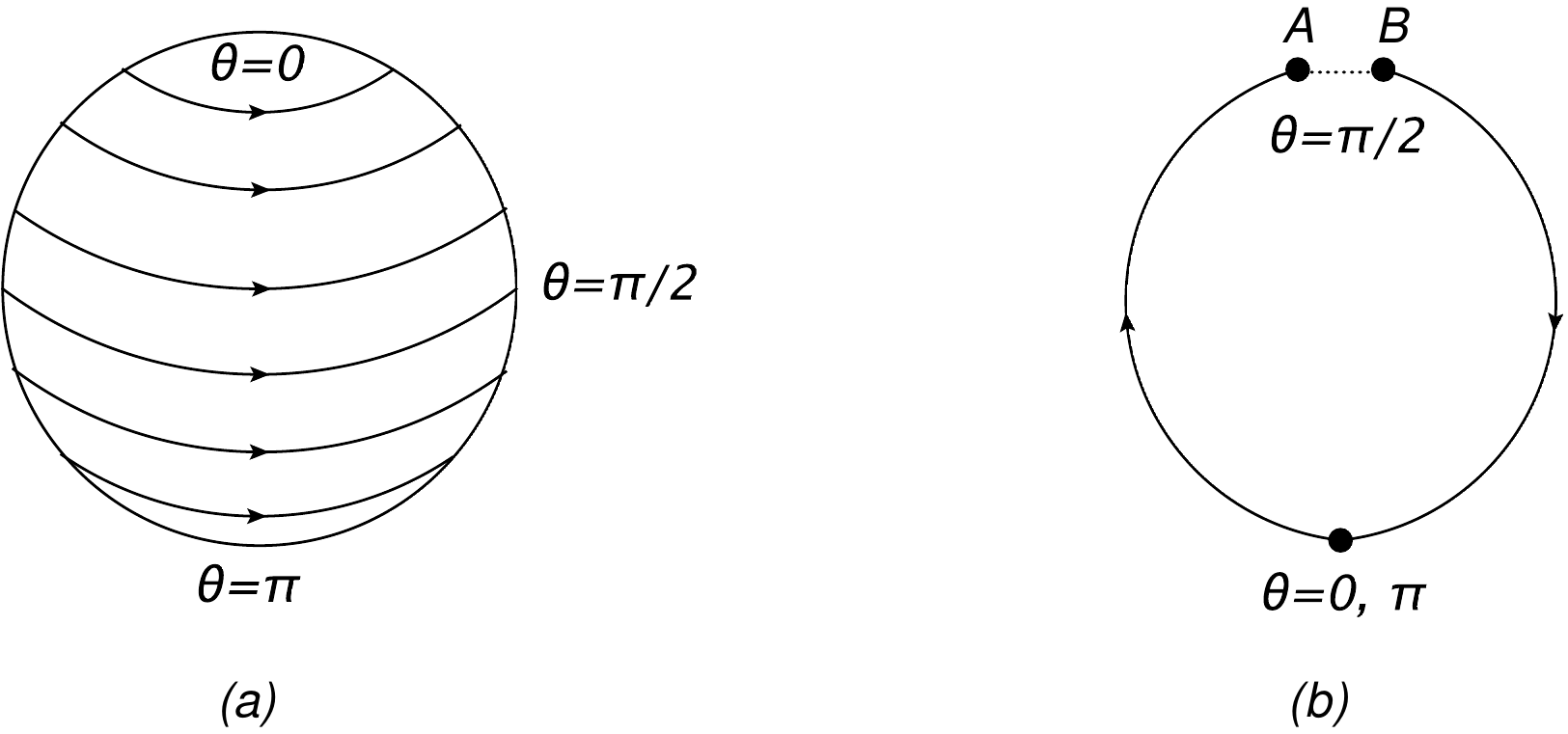}}
\caption{(a) A family of curves for fixed values of the polar angle
  $\theta$, where $\theta=0$ and $\theta=\pi$ correspond to the north
  and south poles respectively; (b) closed loop in the gauge group $G$
  associated with the family of curves.}
\label{fig:curves}
\end{center}
\end{figure}

There is another way to derive the quantisation condition, which makes
the relevant topological aspects clearer, and which will also pave the
way for the non-abelian discussion in the following section. Consider
the family of circles $C(\theta)$ on the 2-sphere at
infinity~\footnote{In fact, any sphere surrounding the origin will do
  e.g. ref.~\cite{Wu:1975es} uses a unit radius.} characterised by
constant values of the polar angle $\theta$. One may associate each
such curve with a group element as follows:
\begin{equation}
U(\theta)=\exp\left[ig\oint_{C(\theta)} dx^\mu A_\mu\right].
\label{Udef} 
\end{equation}
At $\theta=0$, the curve $C(0)$ is an infinitesimally small loop around
the north pole, and thus $U(0)$ corresponds to the identity element of
the gauge group. As $\theta$ increases from $0$ to $\pi$, the curves
gradually sweep over the sphere as shown in figure~\ref{fig:curves}(a),
culminating in an infinitesimally small loop at the south pole, such
that $U(\pi)$ also corresponds to the identity element.
At $\theta=\pi/2$, the definition of the gauge field jumps from the
northern field $A_\mu^N$ to the southern field $A_\mu^S$. We may
depict this as shown in figure~\ref{fig:curves}(b). The curve obtained
in the gauge group as $\theta$ increases from zero proceeds clockwise
from the lower point as shown, reaching point $A$ at
$\theta=\pi/2$. Upon switching to the southern gauge field, there is a
discontinuity and the curve continues from point $B$. However,
eq.~(\ref{Udef}) represents the phase that a charged particle would
experience upon traversing the curve $C(\theta)$. For this to be
single-valued on the equator, it follows that the points $B$ and $A$
should correspond to {\it equivalent} group elements (indicated by the
dotted line). Thus, as $\theta$ varies in the range $[0,\pi]$, a
closed curve is swept out in the gauge group $G$. The set of
topologically non-trivial gauge fields is then in one-to-one
correspondence with the set of topologically inequivalent closed loops
in $G$, namely the possible ways in which one can join the points $A$
and $B$ in figure~\ref{fig:curves}(b). These closed loops are
classified by the {\it first homotopy group} $\pi_1(G)$, where for the
present case one has
\begin{equation}
\pi_1({\mathrm{U(1)}})={\mathbb Z}
\label{pi1U1}
\end{equation}
i.e. the set of integers under addition. This simple result follows
from the fact that the group manifold of $U(1)$ is a circle, such that
any closed loop may wind around this circle an integer number of
times. Put another way, the fact that the two points $A$ and $B$ in
figure~\ref{fig:curves}(b) correspond to equivalent group elements
imposes - via eq.~(\ref{Udef}) - the condition
\begin{equation}
g\oint_{C(\theta)} dx^\mu A_\mu^N
=g\oint_{C(\theta)} dx^\mu A_\mu^S+2\pi n,\quad n\in{\mathbb Z},
\label{condition2}
\end{equation}
which is precisely equivalent to eqs.~(\ref{magflux2},
\ref{Diracquant}).  

\subsection{Non-abelian case}
\label{sec:nonabelian}

In this section, we review the properties of singular point-like
non-abelian magnetic monopoles, where we will follow the presentation
of ref.~\cite{Weinberg:2012pjx} closely. Our ultimate aim is to
identify the relevant quantity that specifies their non-trivial
topology, and to interpret this from the point of view of the double
copy~\footnote{Most of the literature on non-abelian monopoles focuses
  on spontaneously broken theories with additional scalar fields, due
  to the non-singular nature of the solutions. For useful reviews of
  the singular case, see
  e.g. refs.~\cite{Chan:1993rm,Weinberg:2012pjx}.}. First, let us
address the form of a non-abelian monopole field ${\bf A}_\mu=A_\mu^a
{\bf T}^a$.
\subsubsection{Gauge field of a non-abelian monopole}
Assuming a static field, we may choose a gauge such that $A^a_0 =
0$. The remaining components of the gauge field can then be expanded in inverse powers of the radial coordinate:
\begin{equation}
  A^a_i = \f{a^a_i(\theta,\phi)}{r} + \mathcal{O}(r^{-2}) \,,
\end{equation}
where $A^a_i$ and $a^a_i$ belong to the Lie algebra of $G$. This gives rise to a magnetic
field 
\begin{equation}
  B^a_i = -\f{1}{2}\epsilon^{ijk}F^a_{jk} \,,
\end{equation}
where the field strength ${\bf F}_{\mu\nu}=F^a_{\mu\nu}{\bf T}^a$ is
defined by
\begin{equation}\label{Fmunu}
  {\bf F}_{\mu\nu} = \partial_{\mu}{\bf A}_{\nu} - \partial_{\nu}{\bf A}_{\mu} - ig[{\bf A}_{\mu},{\bf A}_{\nu}] \,.
\end{equation}
For a point-like monopole solution we require a magnetic field with
$r^{-2}$ dependence, and therefore the $\mathcal{O}(r^{-2})$ terms in
the potential can be omitted. To avoid any issues arising from the
unavoidable singularity at the origin, we consider the field only for
$r > r_0$, where $r_0$ is the small but non-vanishing radius of a
sphere centred on the monopole. To simplify the analysis, we first
choose a gauge in which $A_r = 0$. This can be done by finding a gauge
transformation such that
\begin{equation}
  {\bf A}_r \to {\bf U} {\bf A}_r {\bf U}^{-1} + {\bf U} \partial_r 
{\bf U}^{-1} = 0 \,.
\end{equation}
A solution to this expression is
\begin{equation}
  {\bf U}^{-1}(r,\theta,\phi) = {\cal P} \exp\lb[\f{i}{g}\int_{r_0}^r \diff r' {\bf A}_r(r',\theta,\phi)\rb] \,,
\end{equation}
where ${\cal P}$ denotes path ordering. This describes an integration
along radial lines, with the lower bound protecting the integral from
the singularity at the origin. A similar procedure can be implemented
for ${\bf A}_{\theta}$ by integrating along lines of constant $r$ and
$\phi$, resulting in a gauge in which ${\bf A}_{\theta}=0$. This
leaves ${\bf A}_{\phi}$ as the only non-zero component of the gauge
field, the form of which may be determined from the equations of motion. The
monopoles considered here are time-independent, so the ${\bf F}_{0i}$
terms in the field strength of eq.~\eqref{Fmunu} vanish. Furthermore,
the assumptions made thus far imply that at large distances ${\bf
  A}_{\phi}$ is independent of $r$. Hence, the only non-vanishing
component of the field strength is
\begin{equation}
  {\bf F}_{\theta\phi} = \partial_{\theta}{\bf A}_{\phi} \,.
\end{equation}
The Yang-Mills field equations in spherical coordinates are
\begin{equation}
  \partial_{\mu}\sqrt{\eta}{\bf F}^{\mu\nu} - ig[{\bf A}_{\mu},\sqrt{\eta}
{\bf F}^{\mu\nu}] = 0 \,,
\end{equation}
where $\eta$ is the absolute value of the determinant of the Minkowski
metric. The field equations give rise to two non-trivial equations of
motion:
\begin{align}
  \partial_{\theta}\sqrt{\eta}{\bf F}^{\theta\phi} &= 0 \,, \label{eom1}\\
  \partial_{\phi}\sqrt{\eta}{\bf F}^{\phi\theta} 
- ig[{\bf A}_{\phi},\sqrt{\eta}{\bf F}^{\phi\theta}] &= 0 \,. \label{eom2} 
\end{align}
Equation~\eqref{eom1} results in
\begin{equation}
  \partial_{\theta}\lb(\f{1}{\sin{\theta}}\partial_{\theta}
{\bf A}_{\phi} \rb) = 0 \,,
\end{equation}
which admits a general solution
\begin{equation}\label{AphiPre}
  {\bf A}_{\phi} = {\bf M}(\phi) + \f{{\bf Q}_M(\phi)}{4\pi}\cos{\theta} \,,
\end{equation}
where ${\bf M}(\phi)$ and ${\bf Q}_M(\phi)$ are matrices in the Lie
algebra of $G$, and the factor of $1/4\pi$ has been included by
convention. As in the abelian case, it is not possible for this
potential to be well-defined for all $\theta$ and we will once again
end up with a Dirac string. Choosing this to lie along the negative
$z$-axis, we require that ${\bf A}_{\phi}$ vanishes at $\theta=0$ to avoid
another singularity at the north pole. This necessitates that
\begin{equation} \label{Mphi}
  {\bf M}(\phi) = -\f{{\bf Q}_M(\phi)}{4\pi} \,.
\end{equation}
Utilising eqs.~\eqref{AphiPre} and~\eqref{Mphi} in the second equation
of motion, eq.~\eqref{eom2}, yields
\begin{equation}
  \partial_{\phi}{\bf Q}_M(\phi) = 0\,,
\end{equation}
and hence ${\bf Q}_M$ is a constant matrix. The general solution for
the gauge field is therefore
\begin{equation}\label{generalA}
  {\bf A}_{\phi} = \f{{\bf Q}_M}{4\pi}(\cos{\theta}-1).
\end{equation}
The importance of this from a double copy perspective is that it
is precisely of the form of eq.~(\ref{singlecopy3}). Thus, faced with
a point-like magnetic monopole in an arbitrary gauge group, one can
always choose a gauge such that the double copy is straightforward,
and will lead to a pure NUT charge in gravity.

\subsubsection{Allowed magnetic charges}
\label{sec:magcharge}

The gauge field of eq.~\eqref{generalA} is defined with a Dirac string
aligned along the negative $z$-axis. We may define this to be a
``northern'' gauge field ${\bf A}_\mu^N$ by analogy with the abelian
case, and define a second potential whose only non-zero component
is
\begin{equation}\label{SgeneralA}
  {\bf A}^S_{\phi} = \f{{\bf Q}_M}{4\pi}(\cos{\theta}+1),
\end{equation}
for use in the southern hemisphere. These two gauge fields are related
by a non-abelian gauge transformation
\begin{equation}\label{equatorGT}
  {\bf A}_\mu^N= {\bf S}(\varphi){\bf A}_\mu^S {\bf S}^{-1}(\varphi)-
\frac{i}{g}{\bf S}(\varphi)\partial_\mu {\bf S}^{-1}(\varphi), 
\end{equation}
where
\begin{equation}
{\bf S}(\varphi)=\exp\left[\frac{ig{\bf Q}_M \varphi}{2\pi}\right].
\label{Sdefnonabel}
\end{equation}
As in section~\ref{sec:abelian}, we may impose the
single-valuedness condition,
\begin{displaymath}
{\bf S}(0)={\bf S}(2\pi),
\end{displaymath}
which leads to a generalised form of the Dirac quantisation
condition:
\begin{equation}\label{QuantCondGen}
  e^{ig{\bf Q}_M} = {\bf I},
\end{equation}
where ${\bf I}$ denotes the identity element in $G$. To analyse this
condition further, one may write the magnetic charge matrix as a
linear combination of the generators ${\bf H}_i$ of the Cartan
  subalgebra of $G$~\cite{Goddard:1976qe}:
\begin{equation}\label{QM}
  {\bf Q}_M = 4\pi \v{k} \cdot \v{H} \,,
\end{equation}
where $\v{k} = (k_1,...,k_r)$ is known as a magnetic weight
vector. Taking this form of the charge matrix in the generalised
quantisation condition, eq.~\eqref{QuantCondGen}, results in
\begin{equation}\label{MagWeightQuant}
  \v{k} \cdot \v{w} = \f{n}{2g}\,, \quad n \in \Z \,,
\end{equation}
where $\v{w}$ is a weight vector of the representation in which the
Cartan generators are expressed. Classifying the possible monopoles in
an arbitrary gauge group $G$ hence reduces to determining which
magnetic weight vectors correspond to physically distinct and stable
monopoles. The magnetic weights are the weights of a Lie group $G^*$ that is dual to $G$. To see this, recall that roots
$\boldsymbol{\alpha}$ and weights $\v{w}$ must always
satisfy~\footnote{For useful reviews of the group theory relevant for the
  present context, see
  e.g. refs.~\cite{Georgi:1999wka,Weinberg:2012pjx}.}
\begin{equation}
  \f{2\v{w} \cdot \boldsymbol{\alpha}}{\boldsymbol{\alpha}^2} = N \,, \quad N \in \Z \,.
\end{equation}
Hence, a possible solution to eq.~\eqref{MagWeightQuant} is to take $\v{k}$ as an element of the root lattice of $G^*$. Recall that the root lattice of a Lie group is the sublattice of the group's weight lattice which contains the root vectors themselves. Thus, the constraint in eq.~\eqref{MagWeightQuant} is satisfied for
\begin{equation}\label{dualRoots}
  \v{k} 
  = \sum_i n_i \boldsymbol{\alpha}^{(i)*} 
  = \sum_i n_i \f{\boldsymbol{\alpha}^{(i)}}{|\boldsymbol{\alpha}^{(i)}|^2} \,,
\end{equation}
where $n_i$ are integers and $\boldsymbol{\alpha}^* =
\boldsymbol{\alpha}/|\boldsymbol{\alpha}|^2$ are the roots of the dual
group $G^*$. We can therefore consider two systems of roots and weights classifying two separate gauge groups, which, following the terminology of ref.~\cite{Goddard:1976qe}, we refer to as the \textit{electric} gauge group $G$ and the \textit{magnetic} gauge group $G^*$. The electric group $G$ has roots $\boldsymbol{\alpha}$ and weights $\v{w}$, while the magnetic group $G^*$ has roots $\boldsymbol{\alpha}^*$ and weights $\v{k}$. The electric weights $\v{w}$ are fixed by the fields present in the theory, and the magnetic weights $\v{k}$ represent the possible magnetic charges. If both groups share the same Lie algebra, then their root vectors will differ only by a rescaling. Furthermore, if $G$ is the universal covering of the algebra then all possible magnetic weights are specified by eq.~\eqref{dualRoots}. In general more solutions will exist. Further examples of electric gauge groups $G$
and their magnetic duals $G^*$ can be found in
ref.~\cite{Goddard:1976qe}. \\

Na\"{i}vely, one would expect an infinite number of possible magnetic
charges, corresponding to arbitrary weights in the dual weight
lattice. However, not all of these correspond to physically distinct
or allowable magnetic monopoles. For example, one may show that
magnetic charges associated with weights ${\bf k}$ and $\v{w}_{{\bf
    k}'}({\bf k})$ are gauge-equivalent, where
\begin{equation}
 \v{w}_{\bf{k}'}(\bf{k})
  = \bf{k} - 2\bf{k}' \f{\bf{k} \cdot \bf{k}'}{\bf{k}' \cdot 
\bf{k}'} \,
\end{equation}
corresponds to a {\it Weyl reflection} in the hyperplane perpendicular
to a third weight ${\bf k}'$. After factoring out Weyl reflections,
the allowed magnetic weights may still live in a number of different {\it sublattices}. Weights which lie within the same sublattice can be connected by an integral sum of roots, while weights in different sublattices cannot. It can be shown that monopoles with higher values of
magnetic charge are dynamically unstable, such that they always decay
to configurations possessing the minimum value of $\tr(Q_M^2)$ within
each sublattice~\cite{Brandt:1979kk}. As the origin of the weight lattice corresponds to the zero magnetic charge configuration, this implies that all monopoles defined by magnetic weights within the same sublattice as the origin will decay to the vacuum state. \\


It is important to note that this instability of monopoles with non-minimal magnetic weights within a given sublattice necessitates a careful specification of the gauge groups. If either the electric or magnetic gauge group is the universal covering group $\tilde{G}$ of the Lie algebra, then the other will be the adjoint group $\tilde{G}/K$, where $K$ is the centre of $\tilde{G}$. Thus if we consider Yang-Mills with $G = \text{SU}(N)$, the magnetic group will be $G^* = \text{SU}(N)/\Z_N$. However, all weights of $\text{SU}(N)/\Z_N$ lie in the same sublattice as the weight at the origin of the lattice. All monopole solutions for $G = \text{SU}(N)$ are therefore dynamically unstable and will reduce to the vacuum solution. This, however, is not the case for $G = \text{SU}(N)/\Z_N$ and $G^* = \text{SU}(N)$, as the $N$ sublattices of the weight lattice of $\text{SU}(N)$ allow for $N-1$ stable monopole solutions. This has an elegant topological interpretation that is directly related to the abelian case, as we now describe.

\subsubsection{Topology of the non-abelian monopole}
\label{sec:topnonabel}

In section~\ref{sec:abelian}, we reviewed the Wu-Yang fibre bundle
interpretation of the Dirac monopole, in which two gauge fields are
patched together, thus generating a gauge configuration with
non-trivial topology. In this section, we describe the generalisation
of this picture to the non-abelian case~\cite{Wu:1975es}. We are again
considering singular monopoles, so that the origin of spacetime is
removed, and spatial slices are then topologically equivalent to
$S^2$. This in turn necessitates at least two coordinate patches,
which we may again choose to be the northern and southern hemispheres
of figure~\ref{fig:patching}(b). Similarly, we may also construct the
family of curves of figure~\ref{fig:curves}, each of which can be
associated with an element of the gauge group,
\begin{equation}
{\bf U}(\theta)={\cal P}\exp\left[ig \oint_{C(\theta)}dx^\mu {\bf A}_\mu
\right],
\label{Wilsongauge}
\end{equation}
where ${\cal P}$ denotes path ordering of the (non-commuting) gauge
fields ${\bf A}_\mu$ along the contour. As $\theta$ varies from $0$ to
$\pi$, this traces out a curve in the electric gauge group $G$,
where the points $\theta=0,\pi$ both correspond to the identity
element. At $\theta=\pi/2$, the gauge field switches from its northern
to southern form, leading once again to the situation shown in
figure~\ref{fig:curves}(b), where $A$ and $B$ must correspond to
equivalent group elements. Denoting the group of such transformations
between $A$ and $B$ by $H$, one may write the general patching
condition between the northern and southern gauge fields as
\begin{equation}
{\cal P}\exp\left[ig \oint_{C}dx^\mu {\bf A}^N_\mu
\right]={\bf U}_H \left\{
{\cal P}\exp\left[ig \oint_{C}dx^\mu {\bf A}^S_\mu
\right]\right\},
\label{patchnonabel}
\end{equation}
where ${\bf U}_H$ constitutes an element of $H$, and $C\equiv
C(\pi/2)$ is the equator of the sphere in
figure~\ref{fig:curves}(a). This is the general condition that encodes
the non-trivial topology of the singular non-abelian monopole. As in
the abelian case, this non-trivial topology is classified by the first
homotopy group of the electric gauge group $\pi_1(G)$, which
follows directly from the fact that $\pi_1(G)$ characterises the
different ways in which one may join the points $A$ and $B$ in
figure~\ref{fig:curves}(b) to form a closed curve. As an example, we
may take an electric group $G=\tilde{G}/K$, in which case
$B$ can be related to $A$ by any element of $K$, the centre of the universal covering
$\tilde{G}$. The relevant first homotopy group is then
\begin{displaymath}
\pi_1\left(\tilde{G}/K \right)=K.
\end{displaymath}
Thus for $G = \text{SU}(N)/\Z_N$, we find $\pi_1(G) = \Z_N$. The identity element
corresponds to monopole configurations that are topologically
equivalent to the vacuum, and hence unstable. This leaves $N-1$ stable
monopoles, consistent with the discussion of the previous section.\\

Note that eq.~(\ref{patchnonabel}) reduces to eq.~(\ref{condition2})
for an abelian electric group $G=\text{U(1)}$. A general
element of $H$ in then given by 
\begin{equation}
{\bf U}_H=e^{2\pi i n},\quad n\in {\mathbb Z},
\label{UHabel}
\end{equation}
from which eq.~(\ref{condition2}) follows. This can in turn be related
to eq.~(\ref{magflux}), which is proportional to the first Chern
number that classifies the non-trivial topology of U(1) principal
bundles. For other gauge groups, however, the patching condition will
not be relatable to the same topological invariant. For example, if
one replaces the gauge group U(1) with $\text{SU}(N)$, the first Chern
number vanishes (due to tracelessness of the generators), but one can
instead classify the topology of solutions using what we may refer to
as the {\it Woodward classes} of ref.~\cite{Woodward}, as we discuss
in appendix~\ref{app:woodward}. The relevance for the double copy is
as follows. Previous studies~\cite{Berman:2018hwd} have suggested that
given characteristic classes in gauge theories may be related to
similar quantities in gravity theories, under the double copy. This
creates a puzzle, in that characteristic classes depend upon the gauge
group, whereas colour information is irrelevant for the gravity side
of the double copy. Here we see that there is in fact no problem, as
it is simply not true that the double copy should apply to individual
characteristic classes. The appropriate quantity that classifies the
``double copiable'' topology of the gauge theory solution is instead
the patching condition of eq.~(\ref{patchnonabel}), whose form is
independent of the gauge group. We will see in the following section
that this indeed has a gravitational counterpart.

\section{Topology of the Taub-NUT solution}
\label{sec:gravity}

The Taub-NUT solution of refs.~\cite{Taub,NUT} is a widely studied
exact solution of GR, whose gravitational field has a rotational
character that does not die off at spatial infinity. As discussed in
section~\ref{sec:review}, the general Taub-NUT solution has a
Schwarzschild-like mass term $M$, and an additional {\it NUT charge}
$N$, where it is the latter that gives rise to the rotational aspect
of the field. Analogies between magnetic monopoles in non-abelian
gauge theories and the Taub-NUT solution have been made many times
before (see e.g. ref.~\cite{Ortin:2004ms} for a review). However,
ref.~\cite{Luna:2015paa} formalised this by pointing out that the
classical double copy implies an exact relationship between dyons and
Taub-NUT, which is true for all values of the radial
coordinate~\footnote{The Taub-NUT solution has also been used to
  examine the double copy of the electromagnetic duality symmetry that
  operates in gauge
  theories~\cite{Alawadhi:2019urr,Banerjee:2019saj,Huang:2019cja}.}. The
double copy is manifest in a particular coordinate system in (2,2)
signature~\cite{Chong:2004hw}, such that the metric assumes the double
Kerr-Schild form of eq.~(\ref{doubleKS}). However, once the double
copy between solutions is known, we may make a coordinate
transformation to more conventional coordinates. To this end, it is
convenient to work with the Taub-NUT line element in spherical polar
coordinates,
\begin{equation}
ds^2=f(r)\left[dt+2N(\cos\theta-1)
d\varphi\right]^2-f^{-1}(r)dr^2-(r^2+N^2)d\Omega^2,
\label{TaubNUTds}
\end{equation}
where 
\begin{equation}
d\Omega^2=d\theta^2+\sin^2\theta d\varphi^2
\label{Omega2}
\end{equation}
is the squared element of solid angle in four spacetime dimensions,
and the additional quantities are given by
\begin{equation}
f(r)=\frac{(r-r_+)(r-r_-)}{r^2+N^2},\quad r_\pm =M\pm \sqrt{M^2+N^2}.
\label{fdef}
\end{equation}
Here $M$ and $N$ are the Schwarzschild-like mass and NUT charge
parameters of eq.~(\ref{doubleKS}) respectively. The case of a pure
NUT charge is given by the limit $M\rightarrow 0$, in which case
eq.~(\ref{TaubNUTds}) reduces to
\begin{equation}
ds^2=\frac{r^2-(\kappa N)^2}{r^2+(\kappa N)^2}\left[dt+2\kappa N(\cos\theta-1)
d\varphi\right]^2-(r^2+(\kappa N)^2)\left[\frac{dr^2}{r^2-(\kappa N)^2}
+d\Omega^2\right],
\label{TaubNUTds2}
\end{equation}
where we have rescaled $N\rightarrow\kappa N$ for later
convenience. The metric has a coordinate singularity at
$\theta=\pi$. This is the so-called {\it Misner string}~\cite{Misner},
and is a direct analogue of the Dirac string for a magnetic
monopole. The known solution to this problem is that one may use two
different coordinate patches, obtained by splitting spatial slices
into a northern and southern hemisphere, as in
figure~\ref{fig:patching}. In the northern hemisphere ($\theta\leq
\pi/2$), the metric of eq.~(\ref{TaubNUTds2}) is well-defined, and one
regards $t$ as a ``northern'' time coordinate $t\equiv t_N$. In the
southern hemisphere, one may transform to a ``southern'' time
coordinate
\begin{equation}
t_N=t_S+4\kappa N\varphi,
\label{tSdef}
\end{equation}
such that the northern and southern line elements are given by
\begin{equation}
ds_{N,S}^2=\frac{r^2-(\kappa N)^2}{r^2+(\kappa N)^2}\left[dt_{N,S}+2\kappa N(\cos\theta\mp 1)
d\varphi\right]^2-(r^2+(\kappa N)^2)\left[\frac{dr^2}{r^2-(\kappa N)^2}
+d\Omega^2\right],
\label{TaubNUTnorthern}
\end{equation}
where the upper (lower) sign corresponds to the northern (southern)
case respectively. One sees that the southern metric is singular at
$\theta=0$, and thus that the Misner string has been moved into the
northern hemisphere. Given that the azimuthal coordinate lies in the
range $0\leq \varphi<2\pi$, eq.~(\ref{tSdef}) (which applies on the
equator on which the two coordinate patches overlap) then implies that
both $t_N$ and $t_S$ must be periodic with period
\begin{equation}
t_0=8\pi \kappa N_0,
\label{t0def}
\end{equation}
where $N_0$ is a basic unit of NUT charge. Indeed, it is known that
this property is required for the spacetime to be fully spherically
symmetric~\cite{HURST196851,Dowker}. The full NUT charge appearing in
eq.~(\ref{TaubNUTnorthern}) is then given by
\begin{equation}
N=nN_0,\quad n\in\mathbb{Z},
\label{NN0}
\end{equation}
where the arbitrary integer $n$ corresponds to the fact that the
transformation between $t_S$ and $t_N$ is defined only up to an
integer multiple of the time period $t_0$~\footnote{From
  eq.~(\ref{t0def}), one may either take $N_0$ as a free parameter, in
  terms of which $t_0$ is fixed, or vice versa. This is analogous to
  the quantisation condition between the electric and magnetic charges
  in abelian gauge theory.}.\\

From the double copy perspective, it is desirable to obtain the above
periodicity condition from a procedure that makes its relationship
with the gauge theory manifest. We may carry this out as
follows. First, a standard result in GR is that there is a {\it time
  holonomy} upon trying to synchronise clocks around a closed
contour. That is, upon traversing a given closed loop $C$, the
difference in time coordinate upon returning to the starting point
satisfies~\cite{Landau:1982dva}
\begin{equation}
|\Delta t|=\oint_C \frac{g_{0i}}{g_{00}}dx^i,
\label{tholonomy}
\end{equation}
where $i\in\{1,2,3\}$ is a spatial index, and the lack of manifest
Lorentz covariance on the right-hand side follows from the explicit
choice of the physical time coordinate on the left-hand side. This
takes a suggestive form if one considers asymptotically large
distances $r\rightarrow\infty$, such that the metrics arising from
eq.~(\ref{TaubNUTnorthern}) take the form
\begin{equation}
g^{N,S}_{\mu\nu}=\eta_{\mu\nu}+\kappa h^{N,S}_{\mu\nu},
\end{equation} 
where the only non-zero components of the northern and southern
gravitons are given by
\begin{equation}
h^{N,S}_{0\varphi}=2 N(\cos\theta \mp 1),\quad
h^{N,S}_{\varphi\varphi}=4\kappa N^2(\cos\theta \mp 1)^2 .
\label{hNS}
\end{equation}
Then the time holonomy of eq.~(\ref{tholonomy}) can be written purely
in terms of the graviton field, as
\begin{equation}
|\Delta t|=\kappa\oint_C h_{0i} d x^i.
\label{tholonomy2}
\end{equation}
Let us now consider the family of curves $C(\theta)$ of
figure~\ref{fig:curves}(a), consisting of circles on the 2-sphere at
spatial infinity. One may associate each such curve with a value of
the time holonomy in eq.~(\ref{tholonomy2}), which from
eqs.~(\ref{hNS}) yields
\begin{equation}
|\Delta t^{N,S}(\theta)|=\kappa\oint_C h^{N,S}_{0\varphi} d \varphi=
2\pi \kappa h^{N,S}_{0\varphi},
\label{DeltNS}
\end{equation}
where we have used the fact that the components
$\{h^{N,S}_{0\varphi}\}$ are independent of $\varphi$. The time shifts
of eq.~(\ref{DeltNS}) form a subgroup of the general group of
diffeomorphisms that act in gravity. One may thus view
eq.~(\ref{DeltNS}) as forming a map from the set of curves in
figure~\ref{fig:curves}(a) to the diffeomorphism group. At $\theta=0$,
the loop $C(0)$ is infinitely small, such that $\Delta t=0$, which
constitutes the identity element of the group. As $\theta$ increases,
a path is traced out in the diffeomorphism group until the equator
$\theta=\pi/2$ is reached. At this point, the graviton field jumps
from its northern to its southern form. Finally, as $\theta\rightarrow
\pi$, the curve $C(\theta)$ becomes infinitely small again, and we are
back at the identity element. This is precisely the picture of
figure~\ref{fig:curves}(b), that we have previously used for elements
of a (non-)abelian gauge group. In order for the gravitons to be
patched on the equator i.e. to correspond to the same physical
solution, the time holonomies arising from $h^N_{\mu\nu}$ and
$h^S_{\mu\nu}$ must be physically equivalent. However, substituting
the explicit results of eqs.~(\ref{hNS}) into
eq.~(\ref{DeltNS}) yields
\begin{equation}
|\Delta t^{S}(\pi/2)|-|\Delta t^N(\pi/2)|=8\pi \kappa N.
\label{Deltres}
\end{equation}
The only way that the time holonomies can coincide is then to impose
periodicity of $t^{N,S}$ with period $t_0$ as in eq.~(\ref{t0def}).\\

There is another way to reach the same conclusion, that makes contact
with Dirac's original argument for the quantisation of electric charge
in an abelian gauge
theory~\cite{Dirac:1931kp}. Reference~\cite{Dowker:1967zz} considered
a gravitational analogue of the magnetic monopole, by considering the
phase experienced by a non-relativistic test particle of mass $m$ that
traverses a given loop $C$ in space, and finding (if $h_{00}=0$) that it is given in the weak field limit by
\begin{equation}
\Phi=\exp\left[i\kappa m\oint_C dx^i h_{0i}\right],
\label{gravphase}
\end{equation}
which is directly related to eq.~(\ref{tholonomy2}). For this phase to
be well-defined on the equator of figure~\ref{fig:curves} where the
two coordinate patches overlap, one must have that the difference in
phases evaluated with the northern and southern graviton fields is a
multiple of $2\pi$, such that 
\begin{equation}
\kappa m\oint_C dx^i \left[h^N_{0i}-h^S_{0i}\right]=2 \pi n,\quad n\in\mathbb{Z}.
\label{gravquant}
\end{equation}
Substituting the results of eqs.~(\ref{hNS}) yields
\begin{equation}
m \kappa N=\frac{n}{4},
\label{mNrel}
\end{equation}
which says that the mass $m$ entering the phase of
eq.~(\ref{gravphase}) is quantised. Reference~\cite{Dowker:1967zz} was
unsure about what to take for this mass. However, its interpretation
becomes clear given the discussion above regarding the periodicity of the
time coordinate. The mass $m$ refers to the energy of a static
wavefunction in the presence of the NUT charge. If the time coordinate
is compact with period $t_0$, this implies that the mass $m$ is
quantised according to
\begin{equation}
\Delta m=\frac{2\pi}{t_0}.
\label{Deltam}
\end{equation}
Equation~(\ref{mNrel}) implies
\begin{equation}
\Delta m=\frac{1}{4 \kappa N_0},
\label{Deltam2}
\end{equation}
so that combining this with eq.~(\ref{Deltam}) yields
eq.~(\ref{t0def}) as required. \\

The description of the quantisation condition in terms of the
gravitational phase experience by a test particle allows one to write
down a patching condition for Taub-NUT that is directly analogous to
the gauge theory condition of eq.~(\ref{patchnonabel}). First, note
that the phase around a loop defines a map from the base spacetime to
the group U(1), so that the curves of figure~\ref{fig:curves}(a) lead
to the description in figure~\ref{fig:curves}(b), where the latter is now
interpreted as a path in the group manifold of U(1). The requirement
that the phases for the northern and southern fields are related by a
multiple of $2\pi$ can be written as
\begin{equation}
\exp\left[i\kappa\oint_C dx^\mu h^N_{0\mu}\right]
=U_H\exp\left[i\kappa\oint_C dx^\mu h^S_{0\mu}\right],
\label{gravpatch}
\end{equation}
where we have used the fact that $h_{00}=0$, and $U_H$ is an element
of the group of transformations that leaves the phase invariant, and
which thus connects the points $A$ and $B$ in
figure~\ref{fig:curves}(b). A general such element is written in
eq.~(\ref{UHabel}), and this description makes clear that the topology
of the Taub-NUT solution is similar to the case of an abelian gauge
theory. That is, the non-trivial topology is classified by maps from
the equator at infinity to U(1), namely by the first homotopy group of
eq.~(\ref{pi1U1}).  A similar conclusion was reached by
ref.~\cite{Bossard:2008sw}, which formally defined the NUT charge
according to the integral~\footnote{Our definition differs slightly to
  that of ref.~\cite{Bossard:2008sw} due to our explicit inclusion of
  factors of $\kappa$.}
\begin{equation}
N=\frac{\kappa}{8\pi}\iint_S \partial_i h_{0j} dx^i\wedge dx^j,
\label{Ndef}
\end{equation}
where $S$ is the 2-sphere at spatial infinity. This can be verified by
separating the field into northern and southern parts related by the
time translations of eq.~(\ref{tSdef}), before applying Stokes'
theorem and using the results of eqs.~(\ref{hNS}). The full NUT charge
is given by an integer multiple of the basic unit $N_0$, according to
eq.~(\ref{NN0}). Indeed, interpreting the integrand of eq.~(\ref{Ndef})
as a field strength, one finds that $N$ is proportional to the first
Chern number, that specifies the non-trivial topology of a U(1)
bundle~\cite{Bossard:2008sw}. The Chern number is integer-valued, and
thus amounts to $n$ in the above construction. Furthermore, the U(1)
fibres in this case correspond to the group of time translations with
a periodic time coordinate, rather than the group of phases
experienced by particles moving along a path. However, the pictures
are ultimately equivalent given that the phase experienced by a
non-relativistic particle is governed directly by the Hamiltonian,
which is conjugate to the time variable.\\

Above, we have seen that one may describe the pure NUT charge in
gravity using one of two descriptions, both of which are
equivalent. That is, one may divide space into two hemispheres and
define the graviton field differently in each, such that the
definitions are related by a diffeomorphism. In the weak field and
non-relativistic limit, one may write a patching condition -
eq.~(\ref{gravpatch}) - that is the precise analogue of its
counterpart in (abelian) gauge theory, and which characterises the
non-trivial topology of the graviton field. The ingredients we have
used to formulate the patching condition of eq.~(\ref{gravpatch}) are
not new. However, they have not previously been analysed from the
point of view of the double copy. The fact that eq.~(\ref{gravpatch})
is a precise counterpart of eq.~(\ref{patchnonabel}) tells us that it
is indeed possible to map global information from gauge theory to
gravity under the double copy, such that the latter becomes more than
merely a local statement. Furthermore, it should be no surprise that
the patching condition on the gravity side takes an abelian-like form,
given that colour structure is stripped off upon taking the double
copy. The patching conditions of eqs.~(\ref{patchnonabel})
and~(\ref{gravpatch}) then form an explicit realisation of
figure~\ref{fig:gaugecopy}, applied to global properties of exact
solutions.\\

It is possible to relate the gauge theory and gravity patching
conditions to previous work in the context of scattering amplitudes,
which we shall do in the following section. Before moving on, however,
it is worth drawing attention to the recent study of
ref.~\cite{Kol:2020ucd}, which provides an alternative way to describe
the Taub-NUT solution by patching together graviton fields analogously
to the Wu-Yang construction~\cite{Wu:1975es}. Instead of relating the
northern and southern fields by a diffeomorphism, the authors instead
use a BMS dual supertranslation~\cite{Godazgar:2018qpq,Kol:2019nkc},
which is defined only at asymptotic infinity, and thus not formally
equivalent to a diffeomorphism. They then argue that a periodic time
coordinate is no longer needed, which avoids the well-known problem
that the Taub-NUT metric gives rise to pathological closed time-like
curves. There are similarities between ref.~\cite{Kol:2020ucd} and the
analysis carried out here, most notably the use of a Wu-Yang-like
formulation for the graviton patching, and also the fact that our
patching condition of eq.~(\ref{gravpatch}) is defined in the weak
field limit, and thus at asymptotic infinity. However, there is a
crucial difference between our ethos and that of
ref.~\cite{Kol:2020ucd}: the double copy for the Taub-NUT solution
applies to the entire classical solutions~\cite{Luna:2015paa}, not
just at asymptotic infinity. Furthermore, the fact that gauge
transformations on the gauge theory side of the correspondence should
be associated with diffeomorphisms on the gravity side is well-known
in the double copy literature on both exact classical solutions and
scattering amplitudes. We have thus used the traditional formulation
of Taub-NUT in this paper, but stress that the relevance (or
otherwise) of ref.~\cite{Kol:2020ucd} deserves further study. It is
also worth emphasising that the results of ref.~\cite{Kol:2020ucd} are
useful for the study of Taub-NUT metrics, independently of the double
copy.

\section{Wilson lines and the double copy}
\label{sec:Wilson}

In the previous sections, we have seen that the non-trivial topology
of a (non)-abelian magnetic monopole can be written in terms of a
certain patching condition -- eq.~(\ref{patchnonabel}) -- that
involves the gauge-covariant phase experienced by a particle moving
around the equator of the 2-sphere at infinity. This corresponds to a
so-called {\it Wilson line} operator, or a {\it Wilson loop} in this
case due to the closed nature of the contour. Such operators appear in
many places in the study of quantum field theory, arising whenever
gauge-dependent information must be compared at different points in
spacetime. What is perhaps less well-known is that Wilson lines have
also been defined in gravity: see e.g.
refs.~\cite{Mandelstam:1962us,Modanese:1993zh,Hamber:2009uz,Brandhuber:2008tf,Donnelly:2015hta}
for (in some cases very) early works on this subject. In recent times,
gravitational Wilson lines have been used to try to set up common
languages between non-abelian gauge theories and
gravity~\cite{Naculich:2011ry,White:2011yy,Miller:2012an,Melville:2013qca,Luna:2016idw},
which is particularly convenient from a double copy point of view. In
this section, we point out how the results of the previous section can
be expressed in terms of Wilson lines, and how this relates to other
examples of the double copy in the literature.\\

Motivated by the gauge theory case, we define a gravitational Wilson
line in terms of the phase experienced by a scalar test particle in a
gravitational field. In the full covariant theory, this must be
proportional to the proper length of a given path $C$, so that we
define
\begin{equation}
\Phi_{\rm grav.}(C)=\exp\left[im\int_C ds \left(g_{\mu\nu} \frac{dx^\mu}{ds} 
\frac{dx^\nu}{ds}\right)^{1/2}\right] 
\label{Phigrav}
\end{equation}
where $x^\mu(s)$ is a parametrisation of the curve, and $m$ the mass
of the test particle. This expression simplifies in the weak field
limit. Writing
\begin{displaymath}
g_{\mu\nu}=\eta_{\mu\nu}+\kappa h_{\mu\nu},
\end{displaymath}
one may expand eq.~(\ref{Phigrav}) in $\kappa$ to obtain
\begin{equation}
\Phi_{\rm grav.}(C)=\exp\left[\frac{i\kappa}{2}\int_C ds \,\frac{dx^\mu}{ds}\,
\frac{dx^\nu}{ds}\,h_{\mu\nu}(x)\right],
\label{Phigrav2}
\end{equation}
where we have ignored an overall normalisation constant that is
independent of $h_{\mu\nu}$ (and that will cancel in any normalised
expectation value of Wilson line operators). We have also absorbed the
mass $m$ into the length parameter $s$. Equation~(\ref{Phigrav2}) is
the Wilson line operator that was considered in
e.g. refs.~\cite{Brandhuber:2008tf,Naculich:2011ry,White:2011yy,Melville:2013qca,Luna:2016idw},
where it was used to analyse properties of scattering
amplitudes~\footnote{Although we considered massive particles above,
  eq.~(\ref{Phigrav2}) can be generalised for massless particles. In
  that case, one must replace the exponent in eq.~(\ref{Phigrav}) with
  the action for a massless point particle, involving an einbein (see
  e.g. ref.~\cite{Green:1987sp}).}. Here we may make contact with
eq.~(\ref{gravphase}) by taking $C$ to be the equator of the 2-sphere
at infinity, and taking $s=mt$, where $t$ is the conventional time
coordinate:
\begin{equation}
\Phi_{\rm grav.}(C)=\exp\left[\frac{i\kappa m}{2}\oint_C dt \left(h_{00}+
2\dot{x}^i h_{0i}+\dot{x}^i\dot{x}^j h_{ij}\right)\right],
\label{WilsonNUT}
\end{equation}
where the dot represents differentiation with respect to $t$. For the
Taub-NUT solution considered in the previous section,
$h_{00}=0$. Then, in the non-relativistic (small velocity) limit,
eq.~(\ref{WilsonNUT}) reduces to eq.~(\ref{gravphase}) as required. \\

There is a very well-defined sense in which the gravitational Wilson
line operator of eq.~(\ref{Phigrav2}) is a double copy of its gauge
theory counterpart
\begin{equation}
\Phi(C)={\cal P}\exp\left[ig \int_C ds \frac{dx^\mu}{ds} {\bf T}^a A^a_\mu(x)
\right],
\label{Phidef}
\end{equation}
where ${\bf T}^a$ is a generator of the gauge group in the appropriate
representation (n.b. this is absorbed into the gauge field in
eq.~(\ref{patchnonabel})), and $s$ a parameter along the curve with
the same mass dimension as in eq.~(\ref{Phigrav2}). To go from
eq.~(\ref{Phidef}) to eq.~(\ref{Phigrav2}), one must replace the
coupling constant with its gravitational counterpart:
\begin{equation}
g\rightarrow \frac{\kappa}{2}
\label{greplace}
\end{equation}
which is precisely the BCJ prescription for scattering
amplitudes~\cite{Bern:2010yg}. Furthermore, one must also strip off
the colour generator ${\bf T}^a$, and replace this with a second
kinematic factor representing the tangent vector to the Wilson line
contour:
\begin{equation}
{\bf T}^a\rightarrow \frac{dx^\nu}{ds}.
\label{Tareplace}
\end{equation}
This mirrors the replacement of colour information by kinematics in
the scattering amplitude double copy. Indeed, there are existing cases
of the amplitude double copy that can be entirely expressed in terms
of Wilson lines, such that the underlying mechanism of the double copy
is precisely the replacements made above. This has not necessarily
been realised in the existing literature, and so it is worthwhile
to briefly review these examples. 

\subsection{The all-order structure of infrared singularities}
\label{sec:IR}

If one dresses an $n$-point scattering amplitude with virtual gluon or
graviton radiation, one encounters {\it infrared (IR) divergences}
associated with the exchanged gauge bosons becoming ``soft''
(i.e. having vanishing 4-momenta)~\footnote{In (non)-abelian gauge
  theories, one also encounters singularities when radiation is
  collinear with the external particles in the interaction, although
  such singularities are absent in
  gravity~\cite{Weinberg:1965nx,Naculich:2011ry,Akhoury:2011kq,Beneke:2012xa}.}. These
singularities have a universal form that factors off from the full
scattering amplitude ${\cal A}$, which has a simple physical
interpretation: soft radiation has an infinite Compton wavelength, and
thus cannot resolve the underlying interaction that produced the $n$
hard particles. One may thus write (see e.g. ref.~\cite{Gardi:2009zv}
for a pedagogical review)
\begin{equation}
{\cal A}={\cal H}\cdot {\cal S},
\label{Afac}
\end{equation}
where ${\cal H}$ is a so-called {\it hard function} that is completely
finite, and ${\cal S}$ a {\it soft function} that collects all the
soft singularities. The latter is known to have an exponential form,
where calculating additional terms in the logarithm of the soft
function amounts to summing up infrared singularities to all orders in
perturbation theory. The structure of the soft function in QCD and QED
is only partially known~\footnote{Interestingly, the soft function of
  QED is known exactly if there are no propagating fermions.} (see
e.g. ref.~\cite{White:2015wha}), and that of gravity is known
exactly. That is, it has now been well-established that the logarithm
of the soft function in gravity terminates at first order in the
gravitational coupling
constant~\cite{Naculich:2011ry,Akhoury:2011kq}. \\

Reference~\cite{Oxburgh:2012zr} used the above properties to present
all-loop order evidence for the BCJ double copy. That is, the authors
showed that one may isolate IR singularities at any given order in
perturbation theory in either QED or QCD, and double copy them to
obtain the known IR singularities of gravity. The somewhat lengthy
analysis used intricate Feynman-diagrammatic arguments, making clear
how the BCJ duality between colour and kinematics could be satisified
at arbitrary loop orders, if terms outside of the soft limit could be
neglected. Furthermore, the authors noted that the IR singluarities of
either QED or QCD both mapped to the same gravitational results, thus
providing an explicit realisation of figure~\ref{fig:gaugecopy}. \\

Here, we wish to point out that the analysis of
ref.~\cite{Oxburgh:2012zr} would have been drastically simpler using
Wilson lines. It is known in (non)-abelian gauge theories that the
soft function can be expressed as a vacuum expectation value of Wilson
line operators, whose contours correspond to the physical trajectories
\begin{displaymath}
x_i^\mu=sp_i^\mu 
\end{displaymath}
of the outgoing hard particles (see e.g. ref.~\cite{White:2015wha}):
\begin{equation}
{\cal S}=\left\langle 0\left| \prod_i \Phi_i\right|0\right\rangle,\quad
\Phi_i\equiv {\cal P}\exp\left[ig{\bf T}^a p_i^\mu \int_0^\infty ds
A^a_\mu\right].
\label{SVEV}
\end{equation}
The physics of this result is that hard particles emitting soft
radiation cannot recoil, and thus can only change by a phase. For this
phase to have the right gauge covariance properties to form part of a
scattering amplitude, it can only be a Wilson line. Armed with the
gravitational Wilson line operator of eq.~(\ref{Phigrav2}), we might
write a similar definition in gravity~\cite{Naculich:2011ry}:
\begin{equation}
{\cal S}_{\rm grav.}=\left\langle 0\left| \prod_i \Phi_{{\rm grav.},i}
\right|0\right\rangle,\quad \Phi_{{\rm grav.}, i}\equiv \exp\left[
  \frac{i\kappa}{2}p_i^\mu p_i^\nu\int_0^\infty ds h_{\mu\nu}\right].
\label{SVEV2}
\end{equation}
The all-order double copy of IR singularities obtained using
diagrammatic arguments then follows simply from the observations of
eqs.~(\ref{greplace}, \ref{Tareplace}), namely that the Wilson line
exponents themselves double copy in a precise way. Furthermore, the
fact that both abelian and non-abelian results both reproduce the same
gravity result is obvious; one strips off the colour generator ${\bf
  T}^a$ in eq.~(\ref{SVEV}), so that it does not matter if one starts
with a QED or QCD Wilson line. \\

There is potentially a rather subtle flaw in the above argument,
namely that for VEVs of Wilson lines to double copy in arbitrary
circumstances, the propagator for the graviton must be written in a
form which manifestly decouples left and right-indices. This is
related to the fact~\cite{Bern:1999ji} that the double copy of pure
gauge theory is not pure gravity, but a theory containing a dilaton
and axion (two-form) field. In any case, the axion and dilaton do not
pose a problem: both are scalar degrees of freedom in four spacetime
dimensions, and by standard power-counting arguments will not
contribute to the structure of leading IR singularities at each order
in perturbation theory~\cite{Oxburgh:2012zr}. Going beyond the leading
soft approximation, one may indeed be sensitive to the additional
matter content (see e.g.~\cite{Plefka:2018dpa} for an interesting
discussion of this point).

\subsection{The Regge limit}
\label{sec:Regge}

\begin{figure}
\begin{center}
\scalebox{0.6}{\includegraphics{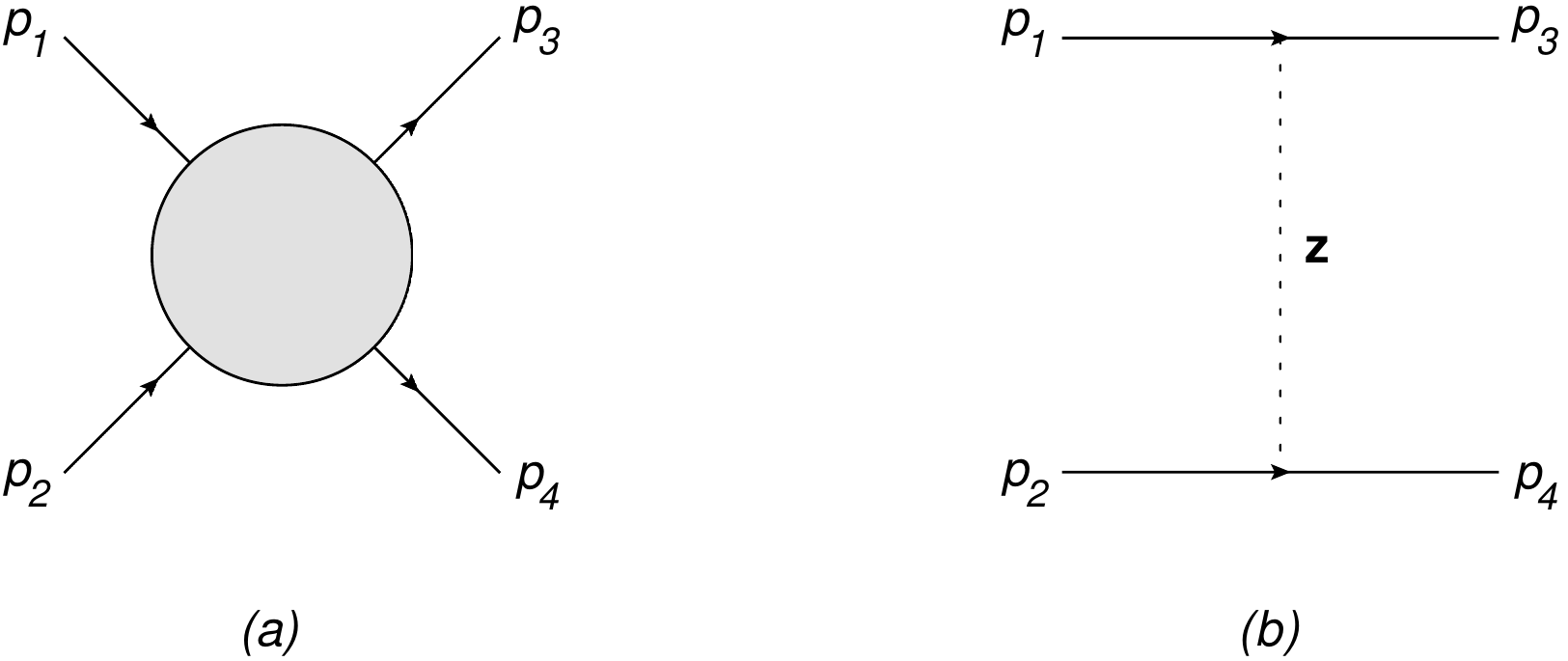}}
\caption{(a) Momentum labels for $2\rightarrow 2$ scattering; (b) In
  the Regge limit, one may model the scattering by two Wilson lines
  separated by a transverse impact parameter $\vec{z}$.}
\label{fig:Regge}
\end{center}
\end{figure}

Another kinematic limit of scattering amplitudes in which all-order
information is obtainable is the high-energy or {\it Regge limit} of
$2\rightarrow 2$ scattering, which can be expressed (for massless
particles) as 
\begin{equation}
s\gg -t,\quad s=(p_1+p_2)^2,\quad t=(p_1-p_3)^2,
\label{Reggedef}
\end{equation}
where we have labelled 4-momenta as in figure~\ref{fig:Regge}(a), and the
Mandelstam invariants $s$ and $t$ constitute the squared centre of
mass energy and momentum transfer respectively.
In the Regge limit of eq.~(\ref{Reggedef}), the particles suffer a
very small deflection owing to the small momentum transfer. Thus, in
position space, they follow approximately straight-line trajectories,
such that the outgoing particles become approximately collinear with
the incoming ones i.e.
\begin{displaymath}
p_3\simeq p_1,\quad p_4\simeq p_2.
\end{displaymath}
The lack of recoil means that the particles can only change by a phase
and, using similar arguments to the previous section, we arrive at the
idea that $2\rightarrow 2$ scattering in the Regge limit can be
described (in position space) by a vacuum expectation value of
Wilson lines separated by a transverse displacement $\vec{z}$, where
the magnitude of the latter is the {\it impact parameter}, or distance
of closest approach:
\begin{equation}
{\cal A}_{s\gg |t|}\rightarrow \left\langle 0\left| \Phi(p_1,0)
\Phi(p_2,\vec{z})\right|0\right\rangle,\quad
\Phi(p,\vec{z})={\cal P}\exp\left[i g_s p^\mu\int_{-\infty}^\infty
ds A_\mu(sp+z)\right].
\label{WilsonRegge}
\end{equation}
This framework was first developed in the classic work of
refs.~\cite{Korchemskaya:1994qp,Korchemskaya:1996je}, which showed
that known properties of the Regge limit in QCD emerge from the Wilson
line approach. Reference~\cite{Melville:2013qca} demonstrated that the
same picture works for gravity if the QCD Wilson lines are replaced
with equivalent ones based on eq.~(\ref{Phigrav2}). The Wilson line
language thus makes the calculations in QCD and gravity look
essentially identical, although the physics in each theory turns out
to be very different. Furthermore, the Wilson line double copy
reproduces previous studies of the Regge limit based on diagrammatic
arguments~\cite{Saotome:2012vy,Vera:2012ds,Johansson:2013nsa,Johansson:2013aca},
where the results overlap. \\

Above, we have seen two examples in which the Wilson line double copy
replacements of eqs.~(\ref{greplace}, \ref{Tareplace}) are directly
related to the BCJ double copy for scattering amplitudes. The fact
that Wilson lines also underly our identification of global
topological information for the magnetic monopole and NUT charge
suggests that they form a useful bridge, relating different
manifestations of the double copy. This also begs the question of what
else can be done with gravitational Wilson lines, an issue which
certainly merits further study.

\section{Conclusion}
\label{sec:conclude}

In this paper, we have examined the double copy for exact classical
solutions, which is normally expressed as a local statement relating
the graviton field to products of gauge fields. We have instead
considered whether it is possible to map {\it global} properties of
solutions between theories, using the magnetic monopole in gauge
theory as a test case. A non-trivial global structure for monopole
solutions is usually characterised by topological invariants that
depend on the field strength. However, these invariants can be
different for different gauge groups, which makes it difficult to see
how to associate this information with a unique gravity counterpart. \\

To clarify these issues, we first reviewed how to express a general
singular monopole solution for arbitrary gauge groups. By a suitable
gauge choice, this can always be expressed as a constant colour matrix
multiplying an abelian-like monopole field, so that the double copy to
gravity is straightforward: one simply removes the colour matrix, and
replaces the monopole with a pure NUT charge, as follows from
ref.~\cite{Luna:2015paa}. This provides an explicit realisation of
figure~\ref{fig:gaugecopy} for exact classical solutions, and extends
the recent SU(2) results of ref.~\cite{Bahjat-Abbas:2020cyb} to
arbitrary gauge groups. \\

We then reviewed the Wu-Yang fibre bundle construction for magnetic
monopoles~\cite{Wu:1975es}, in which one defines separate gauge fields
associated with the northern and southern spatial hemispheres. These
are patched together on the equator, to create a global field that is
free of string-like singularities, but topologically non-trivial. The
topology is classified by the patching condition, which can always be
expressed in terms of Wilson lines representing the phase experienced by
a particle upon being taken around the equator
(eq.~(\ref{patchnonabel})). Next, we showed that a similar
construction can be applied in gravity (see also the recent
ref.~\cite{Kol:2020ucd}), using the known time translations that
remove the Misner string singularity in each coordinate
patch~\cite{Misner}, and which result in a periodic time
coordinate. This treats the NUT solution as a U(1) bundle, where one
may take the fibres to correspond either to time translations (modulo
the time period), or the group of phases experienced by a
non-relativistic particle being taken around a loop. The latter
description maps most cleanly to the Aharonov-Bohm effect in
magnetism~\cite{Dowker:1967zz}, and the resulting patching condition
can then be expressed in terms of gravitational Wilson lines.\\ 

As we argued in section~\ref{sec:Wilson}, the appearance of Wilson
lines makes the double copy particularly natural, given that the
relevant gauge theory and gravity operators are related by simple
replacements of coupling constants, and colour information by
kinematics. We also saw that previous double copy properties of
scattering amplitudes had neat interpretations in terms of Wilson
lines. One of these (the Regge limit) was already
known~\cite{Melville:2013qca}, but the other (the double copy
structure of IR singularities in fixed-angle
scattering~\cite{Oxburgh:2012zr}) has not been previously understood
in this way in the literature. This strongly suggests that there are
other uses of gravitational Wilson lines in a double copy context, and
that they potentially have a key role to play in establishing a
bridge, where appropriate, between the double copy literature on
scattering amplitudes and that relating to classical solutions.\\

We hope that the results of this study provide useful food for thought
in widening the remit of the double copy yet further, including
non-perturbative and / or global aspects.

\section*{Acknowledgments}

We thank Prarit Agarwal, George Barnes, David Berman, Adrian Padellaro
and Costis Papageorgakis for useful discussions. This work has been
supported by the UK Science and Technology Facilities Council (STFC)
Consolidated Grant ST/P000754/1 ``String theory, gauge theory and
duality'', and by the European Union Horizon 2020 research and
innovation programme under the Marie Sk\l{}odowska-Curie grant
agreement No. 764850 ``SAGEX''.

\appendix

\section{Topological classification of SU($N$)/$\mathbb{Z}_N$ bundles}
\label{app:woodward}

In section~\ref{sec:nonabelian}, we reviewed how the non-trivial
topology of a Dirac monopole can be described by the first Chern
class. In the case of other gauge groups, it is not necessarily the
case that a well-known invariant characterises the non-trivial
topology obtained by patching together gauge fields in the northern
and southern spatial hemispheres. However, for SU($N$) groups there is
indeed a known description, which we thought worth pointing out in
this appendix. These results were first obtained by Woodward in
ref.~\cite{Woodward}. \\

As discussed in section~\ref{sec:nonabelian}, we consider magnetic
solutions associated with an electric gauge group
$G=$SU($N$)/${\mathbb Z}_N$, otherwise known as PU($N$). More
formally, we can write the following {\it short exact sequence}
relating the various groups appearing above:
\begin{equation}\label{ses}
    \begin{tikzcd}[row sep=9ex, column sep=9ex] 
        \mathbb{Z}_N \arrow[r, hook, "\text{center}"] & \mathrm{SU}(N) \arrow[r, "\text{quotient}", two heads] & \mathrm{PU}(N)\cong \mathrm{SU}(N) / \mathbb{Z}_N.
    \end{tikzcd}
\end{equation}
Associated with this sequence is a corresponding exact sequence
relating the following homotopy groups:
\begin{equation}
    \begin{aligned}
        \cdots \rightarrow \pi_1\big(\mathbb{Z}_N\big) \rightarrow \pi_1\big(
\mathrm{SU}(N)\big) \rightarrow \pi_1\big(\mathrm{PU}(N)\big) \rightarrow  \pi_0(\mathbb{Z}_N) \rightarrow \pi_0\big(\mathrm{SU}(N)\big) \rightarrow \pi_0\big(\mathrm{PU}(N)\big) \\
        \label{sequence1}
    \end{aligned}
\end{equation}
from which we obtain the isomorphism:
\begin{equation}
    \begin{aligned}
        \pi_1\big(\mathrm{PU}(N)\big)\,&\cong\, \pi_0(\mathbb{Z}_N)=\mathbb{Z}_N.  \\
        \label{pi1PUN}
    \end{aligned}
\end{equation}
The first homotopy group on the left-hand side is what classifies
PU($N$) bundles on $S^2$ base manifolds, as explained previously using
figure~\ref{fig:curves}. Such manifolds arise in our context due to
the fact that all of the magnetic monopoles we consider are singular
at the origin, and thus are defined on ${\mathbb R}^3-\{0\}$, which is
homotopic to $S^2$. Each topologically distinct monopole solution
corresponds to a distinct fibre bundle, and from eq.~(\ref{pi1PUN}),
we see that there are $N$ topologically different monopole solutions
associated with electric group PU($N$), each corresponding to a
different element of the first homotopy group. However, we can also
classify the solutions using the dual description of cohomology
classes, where eq.~(\ref{sequence1}) implies the exact sequence
\begin{equation}
    \begin{aligned}
       \cdots \rightarrow H^1\!\big(S^2,\,\mathrm{SU}(N)\big) \rightarrow H^1\!\big(S^2,\,\mathrm{PU}(N)\big) \rightarrow H^2(S^2,\,\mathbb{Z}_{N})\rightarrow H^2(S^2,\,\mathrm{SU}(N)) \rightarrow \cdots \\
    \end{aligned}
\end{equation}
and hence the isomorphism of cohomology groups
\begin{equation}
    \begin{aligned}
        H^1\!\big(S^2,\,\mathrm{PU}(N)\big)
 &\;\cong\, H^2(S^2,\,\mathbb{Z}_{N})=\mathbb{Z}_N  \\
    \end{aligned}
\end{equation}
We refer to $[\omega]\in H^2(S^2,\,\mathbb{Z}_{N})$ as the $2$nd
\textit{Woodward class} of the PU($N$)-bundle, after
ref.~\cite{Woodward}. For $N=2$ we have PU(2)=SO(3) and the $2$nd
Woodward class reduces to the $2$nd Stiefel-Whitney class. In other
words, the Stiefel-Whitney class replaces the first Chern class as the
relevant characteristic class for monopoles in SU(2) pure gauge
theory. For SU($N$) with $N>2$, one must use the Woodward classes.

\bibliography{refs.bib}

\providecommand{\href}[2]{#2}\begingroup\raggedright\begin{thebibliography}{100}

\bibitem{Bern:2008qj}
Z.~Bern, J.~Carrasco, and H.~Johansson, ``{New Relations for Gauge-Theory
  Amplitudes},'' {\em Phys.Rev.} {\bf D78} (2008) 085011,
\href{http://www.arXiv.org/abs/0805.3993}{{\tt 0805.3993}}.

\bibitem{Bern:2010ue}
Z.~Bern, J.~J.~M. Carrasco, and H.~Johansson, ``{Perturbative Quantum Gravity
  as a Double Copy of Gauge Theory},'' {\em Phys.Rev.Lett.} {\bf 105} (2010)
  061602, \href{http://www.arXiv.org/abs/1004.0476}{{\tt 1004.0476}}.

\bibitem{Bern:2010yg}
Z.~Bern, T.~Dennen, Y.-t. Huang, and M.~Kiermaier, ``{Gravity as the Square of
  Gauge Theory},'' {\em Phys.Rev.} {\bf D82} (2010) 065003,
  \href{http://www.arXiv.org/abs/1004.0693}{{\tt 1004.0693}}.

\bibitem{Bern:1998ug}
Z.~Bern, L.~J. Dixon, D.~Dunbar, M.~Perelstein, and J.~Rozowsky, ``{On the
  relationship between Yang-Mills theory and gravity and its implication for
  ultraviolet divergences},'' {\em Nucl.Phys.} {\bf B530} (1998) 401--456,
\href{http://www.arXiv.org/abs/hep-th/9802162}{{\tt hep-th/9802162}}.

\bibitem{Green:1982sw}
M.~B. Green, J.~H. Schwarz, and L.~Brink, ``{N=4 Yang-Mills and N=8
  Supergravity as Limits of String Theories},'' {\em Nucl.Phys.} {\bf B198}
  (1982)
474--492.

\bibitem{Bern:1997nh}
Z.~Bern, J.~Rozowsky, and B.~Yan, ``{Two loop four gluon amplitudes in N=4
  superYang-Mills},'' {\em Phys.Lett.} {\bf B401} (1997) 273--282,
\href{http://www.arXiv.org/abs/hep-ph/9702424}{{\tt hep-ph/9702424}}.

\bibitem{Carrasco:2011mn}
J.~J. Carrasco and H.~Johansson, ``{Five-Point Amplitudes in N=4
  Super-Yang-Mills Theory and N=8 Supergravity},'' {\em Phys.Rev.} {\bf D85}
  (2012) 025006,
\href{http://www.arXiv.org/abs/1106.4711}{{\tt 1106.4711}}.

\bibitem{Carrasco:2012ca}
J.~J.~M. Carrasco, M.~Chiodaroli, M.~Günaydin, and R.~Roiban, ``{One-loop
  four-point amplitudes in pure and matter-coupled N=4 supergravity},'' {\em
  JHEP} {\bf 1303} (2013) 056,
\href{http://www.arXiv.org/abs/1212.1146}{{\tt 1212.1146}}.

\bibitem{Mafra:2012kh}
C.~R. Mafra and O.~Schlotterer, ``{The Structure of n-Point One-Loop Open
  Superstring Amplitudes},'' {\em JHEP} {\bf 1408} (2014) 099,
\href{http://www.arXiv.org/abs/1203.6215}{{\tt 1203.6215}}.

\bibitem{Boels:2013bi}
R.~H. Boels, R.~S. Isermann, R.~Monteiro, and D.~O'Connell,
  ``{Colour-Kinematics Duality for One-Loop Rational Amplitudes},'' {\em JHEP}
  {\bf 1304} (2013) 107,
\href{http://www.arXiv.org/abs/1301.4165}{{\tt 1301.4165}}.

\bibitem{Bjerrum-Bohr:2013iza}
N.~E.~J. Bjerrum-Bohr, T.~Dennen, R.~Monteiro, and D.~O'Connell, ``{Integrand
  Oxidation and One-Loop Colour-Dual Numerators in N=4 Gauge Theory},'' {\em
  JHEP} {\bf 1307} (2013) 092,
\href{http://www.arXiv.org/abs/1303.2913}{{\tt 1303.2913}}.

\bibitem{Bern:2013yya}
Z.~Bern, S.~Davies, T.~Dennen, Y.-t. Huang, and J.~Nohle, ``{Color-Kinematics
  Duality for Pure Yang-Mills and Gravity at One and Two Loops},''
\href{http://www.arXiv.org/abs/1303.6605}{{\tt 1303.6605}}.

\bibitem{Bern:2013qca}
Z.~Bern, S.~Davies, and T.~Dennen, ``{The Ultraviolet Structure of Half-Maximal
  Supergravity with Matter Multiplets at Two and Three Loops},'' {\em
  Phys.Rev.} {\bf D88} (2013) 065007,
\href{http://www.arXiv.org/abs/1305.4876}{{\tt 1305.4876}}.

\bibitem{Nohle:2013bfa}
J.~Nohle, ``{Color-Kinematics Duality in One-Loop Four-Gluon Amplitudes with
  Matter},''
\href{http://www.arXiv.org/abs/1309.7416}{{\tt 1309.7416}}.

\bibitem{Bern:2013uka}
Z.~Bern, S.~Davies, T.~Dennen, A.~V. Smirnov, and V.~A. Smirnov, ``{Ultraviolet
  Properties of N=4 Supergravity at Four Loops},'' {\em Phys.Rev.Lett.} {\bf
  111} (2013), no.~23, 231302,
\href{http://www.arXiv.org/abs/1309.2498}{{\tt 1309.2498}}.

\bibitem{Naculich:2013xa}
S.~G. Naculich, H.~Nastase, and H.~J. Schnitzer, ``{All-loop infrared-divergent
  behavior of most-subleading-color gauge-theory amplitudes},'' {\em JHEP} {\bf
  1304} (2013) 114,
\href{http://www.arXiv.org/abs/1301.2234}{{\tt 1301.2234}}.

\bibitem{Du:2014uua}
Y.-J. Du, B.~Feng, and C.-H. Fu, ``{Dual-color decompositions at one-loop level
  in Yang-Mills theory},''
\href{http://www.arXiv.org/abs/1402.6805}{{\tt 1402.6805}}.

\bibitem{Mafra:2014gja}
C.~R. Mafra and O.~Schlotterer, ``{Towards one-loop SYM amplitudes from the
  pure spinor BRST cohomology},'' {\em Fortsch.Phys.} {\bf 63} (2015), no.~2,
  105--131,
\href{http://www.arXiv.org/abs/1410.0668}{{\tt 1410.0668}}.

\bibitem{Bern:2014sna}
Z.~Bern, S.~Davies, and T.~Dennen, ``{Enhanced Ultraviolet Cancellations in N =
  5 Supergravity at Four Loop},''
\href{http://www.arXiv.org/abs/1409.3089}{{\tt 1409.3089}}.

\bibitem{Mafra:2015mja}
C.~R. Mafra and O.~Schlotterer, ``{Two-loop five-point amplitudes of super
  Yang-Mills and supergravity in pure spinor superspace},''
\href{http://www.arXiv.org/abs/1505.02746}{{\tt 1505.02746}}.

\bibitem{He:2015wgf}
S.~He, R.~Monteiro, and O.~Schlotterer, ``{String-inspired BCJ numerators for
  one-loop MHV amplitudes},'' {\em JHEP} {\bf 01} (2016) 171,
\href{http://www.arXiv.org/abs/1507.06288}{{\tt 1507.06288}}.

\bibitem{Bern:2015ooa}
Z.~Bern, S.~Davies, and J.~Nohle, ``{Double-Copy Constructions and Unitarity
  Cuts},''
\href{http://www.arXiv.org/abs/1510.03448}{{\tt 1510.03448}}.

\bibitem{Mogull:2015adi}
G.~Mogull and D.~O'Connell, ``{Overcoming Obstacles to Colour-Kinematics
  Duality at Two Loops},'' {\em JHEP} {\bf 12} (2015) 135,
\href{http://www.arXiv.org/abs/1511.06652}{{\tt 1511.06652}}.

\bibitem{Chiodaroli:2015rdg}
M.~Chiodaroli, M.~Gunaydin, H.~Johansson, and R.~Roiban, ``{Spontaneously
  Broken Yang-Mills-Einstein Supergravities as Double Copies},''
\href{http://www.arXiv.org/abs/1511.01740}{{\tt 1511.01740}}.

\bibitem{Bern:2017ucb}
Z.~Bern, J.~J.~M. Carrasco, W.-M. Chen, H.~Johansson, R.~Roiban, and M.~Zeng,
  ``{The Five-Loop Four-Point Integrand of N=8 Supergravity as a Generalized
  Double Copy},''
\href{http://www.arXiv.org/abs/1708.06807}{{\tt 1708.06807}}.

\bibitem{Johansson:2015oia}
H.~Johansson and A.~Ochirov, ``{Color-Kinematics Duality for QCD Amplitudes},''
  {\em JHEP} {\bf 01} (2016) 170,
\href{http://www.arXiv.org/abs/1507.00332}{{\tt 1507.00332}}.

\bibitem{Oxburgh:2012zr}
S.~Oxburgh and C.~White, ``{BCJ duality and the double copy in the soft
  limit},'' {\em JHEP} {\bf 1302} (2013) 127,
\href{http://www.arXiv.org/abs/1210.1110}{{\tt 1210.1110}}.

\bibitem{White:2011yy}
C.~D. White, ``{Factorization Properties of Soft Graviton Amplitudes},'' {\em
  JHEP} {\bf 1105} (2011) 060, \href{http://www.arXiv.org/abs/1103.2981}{{\tt
  1103.2981}}.

\bibitem{Melville:2013qca}
S.~Melville, S.~Naculich, H.~Schnitzer, and C.~White, ``{Wilson line approach
  to gravity in the high energy limit},'' {\em Phys.Rev.} {\bf D89} (2014)
  025009,
\href{http://www.arXiv.org/abs/1306.6019}{{\tt 1306.6019}}.

\bibitem{Luna:2016idw}
A.~Luna, S.~Melville, S.~G. Naculich, and C.~D. White, ``{Next-to-soft
  corrections to high energy scattering in QCD and gravity},'' {\em JHEP} {\bf
  01} (2017) 052,
\href{http://www.arXiv.org/abs/1611.02172}{{\tt 1611.02172}}.

\bibitem{Saotome:2012vy}
R.~Saotome and R.~Akhoury, ``{Relationship Between Gravity and Gauge Scattering
  in the High Energy Limit},'' {\em JHEP} {\bf 1301} (2013) 123,
\href{http://www.arXiv.org/abs/1210.8111}{{\tt 1210.8111}}.

\bibitem{Vera:2012ds}
A.~Sabio~Vera, E.~Serna~Campillo, and M.~A. Vazquez-Mozo, ``{Color-Kinematics
  Duality and the Regge Limit of Inelastic Amplitudes},'' {\em JHEP} {\bf 1304}
  (2013) 086,
\href{http://www.arXiv.org/abs/1212.5103}{{\tt 1212.5103}}.

\bibitem{Johansson:2013nsa}
H.~Johansson, A.~Sabio~Vera, E.~Serna~Campillo, and M.~A. Vázquez-Mozo,
  ``{Color-Kinematics Duality in Multi-Regge Kinematics and Dimensional
  Reduction},'' {\em JHEP} {\bf 1310} (2013) 215,
\href{http://www.arXiv.org/abs/1307.3106}{{\tt 1307.3106}}.

\bibitem{Johansson:2013aca}
H.~Johansson, A.~Sabio~Vera, E.~Serna~Campillo, and M.~A. Vazquez-Mozo,
  ``{Color-kinematics duality and dimensional reduction for graviton emission
  in Regge limit},''
\href{http://www.arXiv.org/abs/1310.1680}{{\tt 1310.1680}}.

\bibitem{Bargheer:2012gv}
T.~Bargheer, S.~He, and T.~McLoughlin, ``{New Relations for Three-Dimensional
  Supersymmetric Scattering Amplitudes},'' {\em Phys.Rev.Lett.} {\bf 108}
  (2012) 231601,
\href{http://www.arXiv.org/abs/1203.0562}{{\tt 1203.0562}}.

\bibitem{Huang:2012wr}
Y.-t. Huang and H.~Johansson, ``{Equivalent D=3 Supergravity Amplitudes from
  Double Copies of Three-Algebra and Two-Algebra Gauge Theories},'' {\em Phys.
  Rev. Lett.} {\bf 110} (2013) 171601,
\href{http://www.arXiv.org/abs/1210.2255}{{\tt 1210.2255}}.

\bibitem{Chen:2013fya}
G.~Chen and Y.-J. Du, ``{Amplitude Relations in Non-linear Sigma Model},'' {\em
  JHEP} {\bf 01} (2014) 061,
\href{http://www.arXiv.org/abs/1311.1133}{{\tt 1311.1133}}.

\bibitem{Chiodaroli:2013upa}
M.~Chiodaroli, Q.~Jin, and R.~Roiban, ``{Color/kinematics duality for general
  abelian orbifolds of N=4 super Yang-Mills theory},'' {\em JHEP} {\bf 01}
  (2014) 152,
\href{http://www.arXiv.org/abs/1311.3600}{{\tt 1311.3600}}.

\bibitem{Johansson:2014zca}
H.~Johansson and A.~Ochirov, ``{Pure Gravities via Color-Kinematics Duality for
  Fundamental Matter},'' {\em JHEP} {\bf 11} (2015) 046,
\href{http://www.arXiv.org/abs/1407.4772}{{\tt 1407.4772}}.

\bibitem{Johansson:2017srf}
H.~Johansson and J.~Nohle, ``{Conformal Gravity from Gauge Theory},''
\href{http://www.arXiv.org/abs/1707.02965}{{\tt 1707.02965}}.

\bibitem{Chiodaroli:2017ehv}
M.~Chiodaroli, M.~Gunaydin, H.~Johansson, and R.~Roiban, ``{Gauged
  Supergravities and Spontaneous Supersymmetry Breaking from the Double Copy
  Construction},'' {\em Phys. Rev. Lett.} {\bf 120} (2018), no.~17, 171601,
\href{http://www.arXiv.org/abs/1710.08796}{{\tt 1710.08796}}.

\bibitem{Chen:2019ywi}
G.~Chen, H.~Johansson, F.~Teng, and T.~Wang, ``{On the kinematic algebra for
  BCJ numerators beyond the MHV sector},''
\href{http://www.arXiv.org/abs/1906.10683}{{\tt 1906.10683}}.

\bibitem{Cheung:2020uts}
C.~Cheung and G.~N. Remmen, ``{Entanglement and the Double Copy},''
\href{http://www.arXiv.org/abs/2002.10470}{{\tt 2002.10470}}.

\bibitem{Monteiro:2014cda}
R.~Monteiro, D.~O'Connell, and C.~D. White, ``{Black holes and the double
  copy},'' {\em JHEP} {\bf 1412} (2014) 056,
\href{http://www.arXiv.org/abs/1410.0239}{{\tt 1410.0239}}.

\bibitem{Luna:2015paa}
A.~Luna, R.~Monteiro, D.~O'Connell, and C.~D. White, ``{The classical double
  copy for Taub-NUT spacetime},'' {\em Phys. Lett.} {\bf B750} (2015) 272--277,
\href{http://www.arXiv.org/abs/1507.01869}{{\tt 1507.01869}}.

\bibitem{Luna:2016due}
A.~Luna, R.~Monteiro, I.~Nicholson, D.~O'Connell, and C.~D. White, ``{The
  double copy: Bremsstrahlung and accelerating black holes},''
\href{http://www.arXiv.org/abs/1603.05737}{{\tt 1603.05737}}.

\bibitem{Goldberger:2016iau}
W.~D. Goldberger and A.~K. Ridgway, ``{Radiation and the classical double copy
  for color charges},'' {\em Phys. Rev.} {\bf D95} (2017), no.~12, 125010,
\href{http://www.arXiv.org/abs/1611.03493}{{\tt 1611.03493}}.

\bibitem{Anastasiou:2014qba}
A.~Anastasiou, L.~Borsten, M.~J. Duff, L.~J. Hughes, and S.~Nagy, ``{Yang-Mills
  origin of gravitational symmetries},'' {\em Phys. Rev. Lett.} {\bf 113}
  (2014), no.~23, 231606,
\href{http://www.arXiv.org/abs/1408.4434}{{\tt 1408.4434}}.

\bibitem{Borsten:2015pla}
L.~Borsten and M.~J. Duff, ``{Gravity as the square of Yang–Mills?},'' {\em
  Phys. Scripta} {\bf 90} (2015) 108012,
\href{http://www.arXiv.org/abs/1602.08267}{{\tt 1602.08267}}.

\bibitem{Anastasiou:2016csv}
A.~Anastasiou, L.~Borsten, M.~J. Duff, M.~J. Hughes, A.~Marrani, S.~Nagy, and
  M.~Zoccali, ``{Twin supergravities from Yang-Mills theory squared},'' {\em
  Phys. Rev.} {\bf D96} (2017), no.~2, 026013,
\href{http://www.arXiv.org/abs/1610.07192}{{\tt 1610.07192}}.

\bibitem{Anastasiou:2017nsz}
A.~Anastasiou, L.~Borsten, M.~J. Duff, A.~Marrani, S.~Nagy, and M.~Zoccali,
  ``{Are all supergravity theories Yang-Mills squared?},''
\href{http://www.arXiv.org/abs/1707.03234}{{\tt 1707.03234}}.

\bibitem{Cardoso:2016ngt}
G.~L. Cardoso, S.~Nagy, and S.~Nampuri, ``{A double copy for $ \mathcal{N}=2 $
  supergravity: a linearised tale told on-shell},'' {\em JHEP} {\bf 10} (2016)
  127,
\href{http://www.arXiv.org/abs/1609.05022}{{\tt 1609.05022}}.

\bibitem{Borsten:2017jpt}
L.~Borsten, ``{On $D=6$, $\mathcal{N}=(2,0)$ and $\mathcal{N}=(4,0)$
  theories},''
\href{http://www.arXiv.org/abs/1708.02573}{{\tt 1708.02573}}.

\bibitem{Anastasiou:2017taf}
A.~Anastasiou, L.~Borsten, M.~J. Duff, A.~Marrani, S.~Nagy, and M.~Zoccali,
  ``{The Mile High Magic Pyramid},''
\newblock 2017.
\newblock
\href{http://www.arXiv.org/abs/1711.08476}{{\tt 1711.08476}}.
\newblock

\bibitem{Anastasiou:2018rdx}
A.~Anastasiou, L.~Borsten, M.~J. Duff, S.~Nagy, and M.~Zoccali, ``{BRST
  squared},''
\href{http://www.arXiv.org/abs/1807.02486}{{\tt 1807.02486}}.

\bibitem{LopesCardoso:2018xes}
G.~Lopes~Cardoso, G.~Inverso, S.~Nagy, and S.~Nampuri, ``{Comments on the
  double copy construction for gravitational theories},'' in {\em {17th
  Hellenic School and Workshops on Elementary Particle Physics and Gravity
  (CORFU2017) Corfu, Greece, September 2-28, 2017}}.
\newblock 2018.
\newblock
\href{http://www.arXiv.org/abs/1803.07670}{{\tt 1803.07670}}.
\newblock

\bibitem{Goldberger:2017frp}
W.~D. Goldberger, S.~G. Prabhu, and J.~O. Thompson, ``{Classical gluon and
  graviton radiation from the bi-adjoint scalar double copy},'' {\em Phys.
  Rev.} {\bf D96} (2017), no.~6, 065009,
\href{http://www.arXiv.org/abs/1705.09263}{{\tt 1705.09263}}.

\bibitem{Goldberger:2017vcg}
W.~D. Goldberger and A.~K. Ridgway, ``{Bound states and the classical double
  copy},'' {\em Phys. Rev.} {\bf D97} (2018), no.~8, 085019,
\href{http://www.arXiv.org/abs/1711.09493}{{\tt 1711.09493}}.

\bibitem{Goldberger:2017ogt}
W.~D. Goldberger, J.~Li, and S.~G. Prabhu, ``{Spinning particles, axion
  radiation, and the classical double copy},'' {\em Phys. Rev.} {\bf D97}
  (2018), no.~10, 105018,
\href{http://www.arXiv.org/abs/1712.09250}{{\tt 1712.09250}}.

\bibitem{Luna:2016hge}
A.~Luna, R.~Monteiro, I.~Nicholson, A.~Ochirov, D.~O'Connell, N.~Westerberg,
  and C.~D. White, ``{Perturbative spacetimes from Yang-Mills theory},'' {\em
  JHEP} {\bf 04} (2017) 069,
\href{http://www.arXiv.org/abs/1611.07508}{{\tt 1611.07508}}.

\bibitem{Luna:2017dtq}
A.~Luna, I.~Nicholson, D.~O'Connell, and C.~D. White, ``{Inelastic Black Hole
  Scattering from Charged Scalar Amplitudes},'' {\em JHEP} {\bf 03} (2018) 044,
\href{http://www.arXiv.org/abs/1711.03901}{{\tt 1711.03901}}.

\bibitem{Shen:2018ebu}
C.-H. Shen, ``{Gravitational Radiation from Color-Kinematics Duality},''
\href{http://www.arXiv.org/abs/1806.07388}{{\tt 1806.07388}}.

\bibitem{Levi:2018nxp}
M.~Levi, ``{Effective Field Theories of Post-Newtonian Gravity},''
\href{http://www.arXiv.org/abs/1807.01699}{{\tt 1807.01699}}.

\bibitem{Plefka:2018dpa}
J.~Plefka, J.~Steinhoff, and W.~Wormsbecher, ``{Effective action of dilaton
  gravity as the classical double copy of Yang-Mills theory},'' {\em Phys.
  Rev.} {\bf D99} (2019), no.~2, 024021,
\href{http://www.arXiv.org/abs/1807.09859}{{\tt 1807.09859}}.

\bibitem{Cheung:2018wkq}
C.~Cheung, I.~Z. Rothstein, and M.~P. Solon, ``{From Scattering Amplitudes to
  Classical Potentials in the Post-Minkowskian Expansion},''
\href{http://www.arXiv.org/abs/1808.02489}{{\tt 1808.02489}}.

\bibitem{Carrillo-Gonzalez:2018pjk}
M.~Carrillo-Gonzalez, R.~Penco, and M.~Trodden, ``{Radiation of scalar modes
  and the classical double copy},''
\href{http://www.arXiv.org/abs/1809.04611}{{\tt 1809.04611}}.

\bibitem{Monteiro:2018xev}
R.~Monteiro, I.~Nicholson, and D.~O'Connell, ``{Spinor-helicity and the
  algebraic classification of higher-dimensional spacetimes},''
\href{http://www.arXiv.org/abs/1809.03906}{{\tt 1809.03906}}.

\bibitem{Plefka:2019hmz}
J.~Plefka, C.~Shi, J.~Steinhoff, and T.~Wang, ``{Breakdown of the classical
  double copy for the effective action of dilaton-gravity at NNLO},''
\href{http://www.arXiv.org/abs/1906.05875}{{\tt 1906.05875}}.

\bibitem{Maybee:2019jus}
B.~Maybee, D.~O'Connell, and J.~Vines, ``{Observables and amplitudes for
  spinning particles and black holes},''
\href{http://www.arXiv.org/abs/1906.09260}{{\tt 1906.09260}}.

\bibitem{Johansson:2019dnu}
H.~Johansson and A.~Ochirov, ``{Double copy for massive quantum particles with
  spin},'' {\em JHEP} {\bf 09} (2019) 040,
\href{http://www.arXiv.org/abs/1906.12292}{{\tt 1906.12292}}.

\bibitem{PV:2019uuv}
A.~PV and A.~Manu, ``{Classical double copy from Color Kinematics duality: A
  proof in the soft limit},''
\href{http://www.arXiv.org/abs/1907.10021}{{\tt 1907.10021}}.

\bibitem{Carrillo-Gonzalez:2019aao}
M.~Carrillo~González, R.~Penco, and M.~Trodden, ``{Shift symmetries, soft
  limits, and the double copy beyond leading order},''
\href{http://www.arXiv.org/abs/1908.07531}{{\tt 1908.07531}}.

\bibitem{Bautista:2019evw}
Y.~F. Bautista and A.~Guevara, ``{On the Double Copy for Spinning Matter},''
\href{http://www.arXiv.org/abs/1908.11349}{{\tt 1908.11349}}.

\bibitem{Moynihan:2019bor}
N.~Moynihan, ``{Kerr-Newman from Minimal Coupling},'' {\em JHEP} {\bf 01}
  (2020) 014,
\href{http://www.arXiv.org/abs/1909.05217}{{\tt 1909.05217}}.

\bibitem{Bah:2019sda}
I.~Bah, R.~Dempsey, and P.~Weck, ``{Kerr-Schild Double Copy and Complex
  Worldlines},''
\href{http://www.arXiv.org/abs/1910.04197}{{\tt 1910.04197}}.

\bibitem{CarrilloGonzalez:2019gof}
M.~Carrillo~González, B.~Melcher, K.~Ratliff, S.~Watson, and C.~D. White,
  ``{The classical double copy in three spacetime dimensions},'' {\em JHEP}
  {\bf 07} (2019) 167,
\href{http://www.arXiv.org/abs/1904.11001}{{\tt 1904.11001}}.

\bibitem{Goldberger:2019xef}
W.~D. Goldberger and J.~Li, ``{Strings, extended objects, and the classical
  double copy},''
\href{http://www.arXiv.org/abs/1912.01650}{{\tt 1912.01650}}.

\bibitem{Kim:2019jwm}
K.~Kim, K.~Lee, R.~Monteiro, I.~Nicholson, and D.~Peinador~Veiga, ``{The
  Classical Double Copy of a Point Charge},''
\href{http://www.arXiv.org/abs/1912.02177}{{\tt 1912.02177}}.

\bibitem{Banerjee:2019saj}
A.~Banerjee, E.~Colgáin, J.~A. Rosabal, and H.~Yavartanoo, ``{Ehlers as EM
  duality in the double copy},''
\href{http://www.arXiv.org/abs/1912.02597}{{\tt 1912.02597}}.

\bibitem{Moynihan:2020gxj}
N.~Moynihan and J.~Murugan, ``{On-Shell Electric-Magnetic Duality and the Dual
  Graviton},''
\href{http://www.arXiv.org/abs/2002.11085}{{\tt 2002.11085}}.

\bibitem{Luna:2018dpt}
A.~Luna, R.~Monteiro, I.~Nicholson, and D.~O'Connell, ``{Type D Spacetimes and
  the Weyl Double Copy},'' {\em Class. Quant. Grav.} {\bf 36} (2019) 065003,
\href{http://www.arXiv.org/abs/1810.08183}{{\tt 1810.08183}}.

\bibitem{White:2016jzc}
C.~D. White, ``{Exact solutions for the biadjoint scalar field},'' {\em Phys.
  Lett.} {\bf B763} (2016) 365--369,
\href{http://www.arXiv.org/abs/1606.04724}{{\tt 1606.04724}}.

\bibitem{DeSmet:2017rve}
P.-J. De~Smet and C.~D. White, ``{Extended solutions for the biadjoint scalar
  field},'' {\em Phys. Lett.} {\bf B775} (2017) 163--167,
\href{http://www.arXiv.org/abs/1708.01103}{{\tt 1708.01103}}.

\bibitem{Bahjat-Abbas:2018vgo}
N.~Bahjat-Abbas, R.~Stark-Muchão, and C.~D. White, ``{Biadjoint wires},'' {\em
  Phys. Lett.} {\bf B788} (2019) 274--279,
\href{http://www.arXiv.org/abs/1810.08118}{{\tt 1810.08118}}.

\bibitem{Bahjat-Abbas:2020cyb}
N.~Bahjat-Abbas, R.~Stark-Muchão, and C.~D. White, ``{Monopoles, shockwaves
  and the classical double copy},''
\href{http://www.arXiv.org/abs/2001.09918}{{\tt 2001.09918}}.

\bibitem{Berman:2018hwd}
D.~S. Berman, E.~Chacón, A.~Luna, and C.~D. White, ``{The self-dual classical
  double copy, and the Eguchi-Hanson instanton},''
\href{http://www.arXiv.org/abs/1809.04063}{{\tt 1809.04063}}.

\bibitem{Taub}
A.~H. Taub, ``Empty space-times admitting a three parameter group of motions,''
  {\em Annals of Mathematics} {\bf 53} (1951), no.~3, pp. 472--490.

\bibitem{NUT}
E.~Newman, L.~Tamburino, and T.~Unti, ``Empty‐space generalization of the
  schwarzschild metric,'' {\em Journal of Mathematical Physics} {\bf 4} (1963),
  no.~7, 915--923.

\bibitem{Bossard:2008sw}
G.~Bossard, H.~Nicolai, and K.~S. Stelle, ``{Gravitational multi-NUT solitons,
  Komar masses and charges},'' {\em Gen. Rel. Grav.} {\bf 41} (2009)
  1367--1379,
\href{http://www.arXiv.org/abs/0809.5218}{{\tt 0809.5218}}.

\bibitem{Stephani:2003tm}
H.~Stephani, D.~Kramer, M.~A. MacCallum, C.~Hoenselaers, and E.~Herlt,
``{Exact solutions of Einstein's field equations},''.

\bibitem{Carrillo-Gonzalez:2017iyj}
M.~Carrillo-González, R.~Penco, and M.~Trodden, ``{The classical double copy
  in maximally symmetric spacetimes},'' {\em JHEP} {\bf 04} (2018) 028,
\href{http://www.arXiv.org/abs/1711.01296}{{\tt 1711.01296}}.

\bibitem{Chong:2004hw}
Z.~Chong, G.~Gibbons, H.~Lu, and C.~Pope, ``{Separability and killing tensors
  in Kerr-Taub-NUT-de sitter metrics in higher dimensions},'' {\em Phys.Lett.}
  {\bf B609} (2005) 124--132,
\href{http://www.arXiv.org/abs/hep-th/0405061}{{\tt hep-th/0405061}}.

\bibitem{Wu:1975es}
T.~T. Wu and C.~N. Yang, ``{Concept of Nonintegrable Phase Factors and Global
  Formulation of Gauge Fields},'' {\em Phys. Rev.} {\bf D12} (1975) 3845--3857.
[,504(1975)].

\bibitem{Weinberg:2012pjx}
E.~J. Weinberg, {\em {Classical solutions in quantum field theory}}.
\newblock Cambridge Monographs on Mathematical Physics. Cambridge University
  Press,
2012.
\newblock

\bibitem{Chan:1993rm}
H.-M. Chan and S.~T. Tsou, ``{Some elementary gauge theory concepts},'' {\em
  World Sci. Lect. Notes Phys.} {\bf 47} (1993)
1--156.

\bibitem{Goddard:1976qe}
P.~Goddard, J.~Nuyts, and D.~I. Olive, ``{Gauge Theories and Magnetic
  Charge},'' {\em Nucl. Phys.} {\bf B125} (1977)
1--28.

\bibitem{Georgi:1999wka}
H.~Georgi, ``{Lie algebras in particle physics},'' {\em Front. Phys.} {\bf 54}
  (1999)
1--320.

\bibitem{Brandt:1979kk}
R.~A. Brandt and F.~Neri, ``{Stability Analysis for Singular Nonabelian
  Magnetic Monopoles},'' {\em Nucl. Phys.} {\bf B161} (1979)
253--282.

\bibitem{Woodward}
L.~M. Woodward, ``The classification of principal pun-bundles over a
  4-complex,'' {\em Journal of the London Mathematical Society} {\bf s2-25}
  (1982), no.~3, 513--524,
  \href{http://www.arXiv.org/abs/https://londmathsoc.onlinelibrary.wiley.com/doi/pdf/10.1112/jlms/s2-25.3.513}{{\tt
  https://londmathsoc.onlinelibrary.wiley.com/doi/pdf/10.1112/jlms/s2-25.3.513}}.

\bibitem{Ortin:2004ms}
T.~Ort\'{i}n, {\em Gravity and Strings}.
\newblock Cambridge University Press, 2004.
\newblock Cambridge Books Online.

\bibitem{Alawadhi:2019urr}
R.~Alawadhi, D.~S. Berman, B.~Spence, and D.~Peinador~Veiga, ``{S-duality and
  the Double Copy},''
\href{http://www.arXiv.org/abs/1911.06797}{{\tt 1911.06797}}.

\bibitem{Huang:2019cja}
Y.-T. Huang, U.~Kol, and D.~O'Connell, ``{The Double Copy of Electric-Magnetic
  Duality},''
\href{http://www.arXiv.org/abs/1911.06318}{{\tt 1911.06318}}.

\bibitem{Misner}
C.~W. Misner, ``The flatter regions of newman, unti, and tamburino's
  generalized schwarzschild space,'' {\em Journal of Mathematical Physics} {\bf
  4} (1963), no.~7, 924--937.

\bibitem{HURST196851}
C.~Hurst, ``Charge quantization and nonintegrable lie algebras,'' {\em Annals
  of Physics} {\bf 50} (1968), no.~1, 51 -- 75.

\bibitem{Dowker}
J.~Dowker, ``The {NUT} solution as a gravitational dyon,'' {\em General
  Relativity and Gravitation} {\bf 5} (1974), no.~5, 603--613.

\bibitem{Landau:1982dva}
L.~D. Landau and E.~M. Lifschits, {\em {The Classical Theory of Fields}},
  vol.~Volume 2 of {\em Course of Theoretical Physics}.
\newblock Pergamon Press, Oxford,
1975.
\newblock

\bibitem{Dirac:1931kp}
P.~A.~M. Dirac, ``{Quantized Singularities in the Electromagnetic Field},''
  {\em Proc. Roy. Soc. Lond.} {\bf A133} (1931) 60--72.
[,278(1931)].

\bibitem{Dowker:1967zz}
J.~S. Dowker and J.~A. Roche, ``{The Gravitational analogues of magnetic
  monopoles},'' {\em Proc. Phys. Soc.} {\bf 92} (1967)
1--8.

\bibitem{Kol:2020ucd}
U.~Kol and M.~Porrati, ``{Gravitational Wu-Yang Monopoles},''
\href{http://www.arXiv.org/abs/2003.09054}{{\tt 2003.09054}}.

\bibitem{Godazgar:2018qpq}
H.~Godazgar, M.~Godazgar, and C.~N. Pope, ``{New dual gravitational charges},''
  {\em Phys. Rev.} {\bf D99} (2019), no.~2, 024013,
\href{http://www.arXiv.org/abs/1812.01641}{{\tt 1812.01641}}.

\bibitem{Kol:2019nkc}
U.~Kol and M.~Porrati, ``{Properties of Dual Supertranslation Charges in
  Asymptotically Flat Spacetimes},'' {\em Phys. Rev.} {\bf D100} (2019), no.~4,
  046019,
\href{http://www.arXiv.org/abs/1907.00990}{{\tt 1907.00990}}.

\bibitem{Mandelstam:1962us}
S.~Mandelstam, ``{Quantization of the gravitational field},'' {\em Annals
  Phys.} {\bf 19} (1962)
25--66.

\bibitem{Modanese:1993zh}
G.~Modanese, ``{Wilson loops in four-dimensional quantum gravity},'' {\em Phys.
  Rev.} {\bf D49} (1994) 6534--6542,
\href{http://www.arXiv.org/abs/hep-th/9307148}{{\tt hep-th/9307148}}.

\bibitem{Hamber:2009uz}
H.~W. Hamber and R.~M. Williams, ``{Gravitational Wilson Loop in Discrete
  Quantum Gravity},'' {\em Phys. Rev.} {\bf D81} (2010) 084048,
\href{http://www.arXiv.org/abs/0907.2652}{{\tt 0907.2652}}.

\bibitem{Brandhuber:2008tf}
A.~Brandhuber, P.~Heslop, A.~Nasti, B.~Spence, and G.~Travaglini, ``{Four-point
  Amplitudes in N=8 Supergravity and Wilson Loops},'' {\em Nucl.Phys.} {\bf
  B807} (2009) 290--314,
\href{http://www.arXiv.org/abs/0805.2763}{{\tt 0805.2763}}.

\bibitem{Donnelly:2015hta}
W.~Donnelly and S.~B. Giddings, ``{Diffeomorphism-invariant observables and
  their nonlocal algebra},'' {\em Phys. Rev.} {\bf D93} (2016), no.~2, 024030,
  \href{http://www.arXiv.org/abs/1507.07921}{{\tt 1507.07921}}.
[Erratum: Phys. Rev.D94,no.2,029903(2016)].

\bibitem{Naculich:2011ry}
S.~G. Naculich and H.~J. Schnitzer, ``{Eikonal methods applied to gravitational
  scattering amplitudes},'' {\em JHEP} {\bf 1105} (2011) 087,
\href{http://www.arXiv.org/abs/1101.1524}{{\tt 1101.1524}}.

\bibitem{Miller:2012an}
D.~Miller and C.~White, ``{The Gravitational cusp anomalous dimension from AdS
  space},'' {\em Phys.Rev.} {\bf D85} (2012) 104034,
\href{http://www.arXiv.org/abs/1201.2358}{{\tt 1201.2358}}.

\bibitem{Green:1987sp}
M.~B. Green, J.~H. Schwarz, and E.~Witten, {\em {SUPERSTRING THEORY. VOL. 1:
  INTRODUCTION}}.
\newblock Cambridge Monographs on Mathematical Physics.
1988.
\newblock

\bibitem{Weinberg:1965nx}
S.~Weinberg, ``{Infrared photons and gravitons},'' {\em Phys.Rev.} {\bf 140}
  (1965)
B516--B524.

\bibitem{Akhoury:2011kq}
R.~Akhoury, S.~Ryo, and G.~Sterman, ``{Collinear and Soft Divergences in
  Perturbative Quantum Gravity},'' {\em Phys.Rev.} {\bf D84} (2011) 104040,
\href{http://www.arXiv.org/abs/1109.0270}{{\tt 1109.0270}}.

\bibitem{Beneke:2012xa}
M.~Beneke and G.~Kirilin, ``{Soft-collinear gravity},'' {\em JHEP} {\bf 1209}
  (2012) 066,
\href{http://www.arXiv.org/abs/1207.4926}{{\tt 1207.4926}}.

\bibitem{Gardi:2009zv}
E.~Gardi and L.~Magnea, ``{Infrared singularities in QCD amplitudes},'' {\em
  Nuovo Cim.} {\bf 032C} (2009) 137--157,
\href{http://www.arXiv.org/abs/0908.3273}{{\tt 0908.3273}}.

\bibitem{White:2015wha}
C.~D. White, ``{An Introduction to Webs},'' {\em J. Phys.} {\bf G43} (2016),
  no.~3, 033002,
\href{http://www.arXiv.org/abs/1507.02167}{{\tt 1507.02167}}.

\bibitem{Bern:1999ji}
Z.~Bern and A.~K. Grant, ``{Perturbative gravity from QCD amplitudes},'' {\em
  Phys. Lett.} {\bf B457} (1999) 23--32,
\href{http://www.arXiv.org/abs/hep-th/9904026}{{\tt hep-th/9904026}}.

\bibitem{Korchemskaya:1994qp}
I.~Korchemskaya and G.~Korchemsky, ``{High-energy scattering in QCD and cross
  singularities of Wilson loops},'' {\em Nucl.Phys.} {\bf B437} (1995)
  127--162,
\href{http://www.arXiv.org/abs/hep-ph/9409446}{{\tt hep-ph/9409446}}.

\bibitem{Korchemskaya:1996je}
I.~A. Korchemskaya and G.~P. Korchemsky, ``{Evolution equation for gluon Regge
  trajectory},'' {\em Phys. Lett.} {\bf B387} (1996) 346--354,
\href{http://www.arXiv.org/abs/hep-ph/9607229}{{\tt hep-ph/9607229}}.

\end{thebibliography}\endgroup
\end{document}